\documentclass[review,3p]{elsarticle}

\usepackage{lineno,hyperref}
\modulolinenumbers[1]

\usepackage{amsmath}
\usepackage{subfloat}
\usepackage{caption}
\usepackage{subfigure}

\usepackage{ragged2e} 
\usepackage{graphicx}
\usepackage{caption}
\usepackage{afterpage}
\usepackage[utf8]{inputenc}
\usepackage{array}
\usepackage[T1]{fontenc}
\usepackage{amssymb}
\usepackage{float}
\usepackage{bm}

\journal{International Journal of Human - Computer Studies}

\biboptions{authoryear}

\begin{document}

\begin{frontmatter}

\title{Comparing Hand Gestures and a Gamepad Interface for Locomotion in Virtual Environments\tnoteref{t1}}

\tnotetext[t1]{©2022. This manuscript version is made available under the CC-BY-NC-ND 4.0 license https://creativecommons.org/licenses/by-nc-nd/4.0/}

\author[mymainaddress]{Jingbo Zhao\corref{mycorrespondingauthor}}
\cortext[mycorrespondingauthor]{Corresponding author}
\address[mymainaddress]{College of Information and Electrical Engineering, China Agricultural University, No.17 Tsinghua East Road, Beijing, 100083, China}
\ead{zhao.jingbo@cau.edu.cn}

\author[mymainaddress]{Ruize An}
\author[mymainaddress]{Ruolin Xu}
\author[mymainaddress]{Banghao Lin}

\begin{abstract}
Hand gesture is a new and promising interface for locomotion in virtual environments. While several previous studies have proposed different hand gestures for virtual locomotion, little is known about their differences in terms of performance and user preference in virtual locomotion tasks. In the present paper, we presented three different hand gesture interfaces and their algorithms for locomotion, which are called the Finger Distance gesture, the Finger Number gesture and the Finger Tapping gesture. These gestures were inspired by previous studies of gesture-based locomotion interfaces and are typical gestures that people are familiar with in their daily lives. Implementing these hand gesture interfaces in the present study enabled us to systematically compare the differences between these gestures. In addition, to compare the usability of these gestures to locomotion interfaces using gamepads, we also designed and implemented a gamepad interface based on the Xbox One controller. We conducted empirical studies to compare these four interfaces through two virtual locomotion tasks. A desktop setup was used instead of sharing a head-mounted display among participants due to the concern of the Covid-19 situation. Through these tasks, we assessed the performance and user preference of these interfaces on speed control and waypoints navigation. Results showed that user preference and performance of the Finger Distance gesture were close to that of the gamepad interface. The Finger Number gesture also had close performance and user preference to that of the Finger Distance gesture. Our study demonstrates that the Finger Distance gesture and the Finger Number gesture are very promising interfaces for virtual locomotion. We also discuss that the Finger Tapping gesture needs further improvements before it can be used for virtual walking.
\end{abstract}


\end{frontmatter}


\section{Introduction}

Locomotion in virtual environments is a hotspot for virtual reality (VR) research. Implementing intuitive and natural locomotion interfaces is an important factor for consideration when designing a new VR environment and system. Low-cost optical tracking devices, such as the Kinect sensor and the Leap Motion sensor, for body and hand skeleton tracking have resulted in many methods for body gesture   \citep{lun_survey_2015} and hand gesture recognition \citep{bachmann_review_2018}. Ample opportunities have become available for VR researchers to investigate the utility of hand gestures for virtual locomotion.

One major advantage of these low-cost optical sensors is that they do not require attaching markers or other tracking devices on the body parts to be tracked while providing tracking information that is sufficiently accurate for interaction in VR. Compared to locomotion interfaces using gamepads, hand gesture interfaces do not require users to hold devices in their hands, hence more sophisticated gestures can be recognized and are generally more flexible compared to VR systems using controllers in terms of gesture recognition (but implementation of force feedback without holding controllers or wearing exoskeletons remains a challenging problem). On the other hand, if we compare virtual locomotion interfaces using hand gestures to interfaces using body tracking with the Kinect sensor or other full-body optical tracking systems, hand gesture interfaces do not involve large-scale body movements. This provides users with an advantage that hand gesture interfaces can be operated in places that are limited in physical space. Such scenarios typically include personal gaming, virtual classrooms and teleoperation, \textit{etc.}, in which large-scale tracking may not be available due to constrained physical space or other reasons. Combined with hand gesture recognition methods for other types of interaction, different interaction activities will be made possible in a VR system. More recent VR headsets, such as the Oculus Quest 2 and the Microsoft Hololens 2, have integrated hand-tracking modules and algorithms. This makes hand gesture locomotion interfaces and other interactions using hand gestures more accessible as additional hand-tracking hardware is not required when using these headsets.

Some previous studies on locomotion using hand gestures used different geometric features of hands for gesture recognition
\citep{cardoso_comparison_2016,zhang_double_2017,cisse_user_2020}. In such designs, each hand shape with a set of geometric features corresponds to a specific walking speed. By designing a set of hand gestures with different hand shapes, one can achieve variable walking speed with a set of hand gestures tracked by optical devices (\textit{e.g.} the Leap Motion sensor). Another approach was to use the distance between the index finger and the centre of a tracked hand and map the distance to locomotion speed \citep{huang_design_2019}. In addition, methods that simulate bipedal walking using two fingers were designed based on multi-touch pads \citep{kim_finger_2008,yan_let_2016}. Although previous hand gesture interfaces have been assessed on a case-by-case basis, there lacks a comparison among different hand gesture interfaces to reveal their differences in terms of performance and user preference. In addition, it is also uncertain how these hand gesture interfaces can be compared to locomotion interfaces using gamepads. 

To fill these research gaps, we presented three hand gesture interfaces and their algorithms, called the Finger Distance gesture, the Finger Number gesture and the Finger Tapping gesture, which were inspired by previous studies \citep{kim_finger_2008,huang_design_2019,cardoso_comparison_2016,zhao_learning_2016}. These are typical gestures that people are familiar with in their daily lives or people can naturally come up with when they intend to make hand gestures. For comparison, we also designed and implemented a gamepad interface based on the Xbox One controller. We compared these four interfaces using two virtual locomotion tasks, which are called the target pursuit task and the waypoints navigation task. The first task evaluated the performance of the gestures in terms of their usability in speed control while the second task focused on waypoints navigation as a more general locomotion task with direction control introduced. We also respectively assessed user preference in these tasks through a subjective user interface questionnaire. Due to the current Covid-19 situation, we opted to use a desktop setup to conduct our experiments as sharing headsets among participants is a health concern \citep{steed_evaluating_2020} and the procedure for sanitizing VR headsets has not been established at the moment.

The goal of the present study was to compare the differences of these four interfaces in terms of their performance and user preference in virtual locomotion tasks.
The main contributions of the study are two-fold:

(1) We present three hand gesture interfaces and their algorithms for virtual locomotion.

(2) We systematically evaluate these three hand gesture interfaces and a gamepad interface using two virtual locomotion tasks and demonstrate their performance and user preference.

The rest of the paper is organized as follows: Section 2 discusses related work; Section 3 presents hand gesture interfaces, a user interface questionnaire and VR hardware and software for experiments; Section 4 presents the design and results of two VR experiments that evaluated four interfaces; Section 5 provides discussion and Section 6 concludes the study. 

\section{Related Work}
In this section, we review previous work on virtual locomotion interfaces using hand gestures. Methods using hand gestures for virtual locomotion can be generally divided into two categories: gestures based on touch pads and mid-air methods based on hand-tracking sensors.

For example, \citet{kim_finger_2008} proposed a hand gesture locomotion technique called Finger Walking in Place (FWIP) using a multi-touch pad, which belongs to the first category. The gesture required users to use two fingers: the index finger and the middle finger to move back and forth on a touchpad to simulate natural walking using two legs. An experiment was conducted to evaluate the method using a locomotion task that asked participants to pass a few waypoints. A joystick interface was also introduced as the control interface for comparison. Subjective factors, including satisfaction, fastness, easiness and tiredness, were evaluated but no objective parameters were analysed in the study. Results showed that participants preferred using the joystick interfaces as it offered continuous control mechanism.

Similarly, \citet{yan_let_2016} also proposed locomotion methods based on a multi-touch pad. These methods include the walking gesture, the segway gesture and the surfing gesture, which allowed users to travel at different modes and speeds with different gestures. The walking gesture, as the name suggests, is a method that allows users to use two fingers to simulate bipedal walking. Walking speed is controlled by the length and the frequency of each step made with fingers. Turning is controlled by a rate-based method that depends on the pressure pressed on the left or right side of the pad. The segway gesture is a method that allows users to use two fingers to be placed on the pad horizontally to simulate riding a segway scooter. Locomotion speed is determined by forward and backward pressure of the two fingers and turning is calculated by the pressure difference between two fingers. The surfing gesture requires users to place their thumb and index fingers in a vertical line on the pad. The vertical distance between these two fingers determines the locomotion speed; The horizontal distance between these fingers is used to control the turning rate; and the pressure difference between these two fingers controls the pitch angle. For each proposed gesture, there was a different locomotion task to compare the interface with the gamepad interface. Both objective and subjective factors were studied. Authors concluded that the gamepad interface was more time-efficient while gesture interfaces based on a multi-touch pad had similar quality compared to the gamepad interface. In addition, switching between different travel modes was faster on the multi-touch pad.

Recent methods using hand gestures for virtual locomotion mainly focused on mid-air gestures based on hand-tracking sensors, such as the Leap Motion sensor.

\citet{cardoso_comparison_2016} proposed a method using the number of extended fingers to control locomotion speed. Opening and closing both hands were used to start and stop locomotion. The proposed method was compared to a gaze-based locomotion approach using a head-mounted display and a gamepad interface. Results showed that the gamepad interface was better than two other interfaces. However, details of implementation of the gesture recognition algorithms of the hand gesture interface were not mentioned in their paper and there was only one set of gestures evaluated in the study.

\citet{huang_design_2019} proposed a technique to control locomotion speed based the distance between the tips of their index finger and the centre of a tracked hand. The proposed method was evaluated on a linear locomotion task, in which participants were asked to walk to targets at different distances as quickly as possible. Their study experimented with different combinations of parameters of the algorithm but the method was not compared with other locomotion interfaces using hand gestures.

More recently, \citet{cisse_user_2020} proposed a method to design hand gestures by inviting professionals in architectural designs, with exposure to VR in professional capacity, to elicit their preferred hand gestures for locomotion. In total, sixty-four gestures were elicited from six professionals. These gestures were categorized into eight classes, which included moving forward, moving backward and moving up a floor, \textit{etc}. To evaluate and select a set of efficient gestures, twelve university students were invited to evaluate different combinations of gestures for virtual locomotion. Factors, including task completion time and intuitiveness, \textit{etc}., were included in the study. However, gamepad interfaces were not included for comparison. The designed gestures were static gestures, which potentially lack information that can be given by finger movements to control locomotion speed.

Another recent work by \citet{caggianese_freehand-steering_2020} proposed three different methods to control locomotion direction: (1) using the pointing direction of a finger; (2) using palm orientation; and (3) using head orientation tracked by a VR headset. However, speed control mechanism was not studied in this research. Initiation and termination of locomotion were controlled by extending and closing a tracked hand. To evaluate three different steering control methods, objective factors and subjective factors were considered. A limitation with the study was that locomotion speed was not considered. Thus, directional control mechanisms need to be combined with speed control methods to further evaluate their performance.

\citet{schafer_controlling_2021} compared four different static hand gestures for teleportation-based virtual locomotion using a single hand or both hands. Their results showed that all proposed methods are viable options for virtual locomotion. They also recommend adding these methods in VR systems and letting users choose their preferred methods. A similar technique was also presented by \citet{bozgeyikli_point_2016} that allowed users to point and teleport to a target location. Since these are teleportation techniques, control of locomotion speed is not considered.

Additional techniques for locomotion involving hand tracking include  pushing or tapping \citep{ferracani_locomotion_2016}, using the translation and the orientation of a tracked non-dominant hand (fist) to control walking speed and direction \citep{tomberlin_gauntlet_2017}, tracking controller motions \citep{pai_pinchmove_2018, rantala_comparison_2021}, and TriggerWalking \citep{sarupuri_triggerwalking_2017,sarupuri_evaluating_2017}, which uses triggers on controllers to simulate bipedal walking. 

To conclude, current trends to use hand gestures for virtual locomotion are primarily mid-air approaches. Using static mid-air hand gestures without considering finger movements does not allow the control of locomotion speed. Thus, finger movements need to be taken into account when designing mid-air gestures. In terms of evaluation, there was no comparison made among different mid-air hand gesture interfaces and the gamepad interfaces for virtual locomotion on speed control and waypoints navigation. Thus, our study presented three mid-air hand gesture interfaces based on previous studies with a gamepad interface. We also compared their performance and user preference through two virtual locomotion tasks, which respectively evaluated their usability in these scenarios.

\section{Methods}
\subsection{Finger Distance Gesture}
\begin{figure}[!ht]
    \centering
        \includegraphics[width=0.3\textwidth]{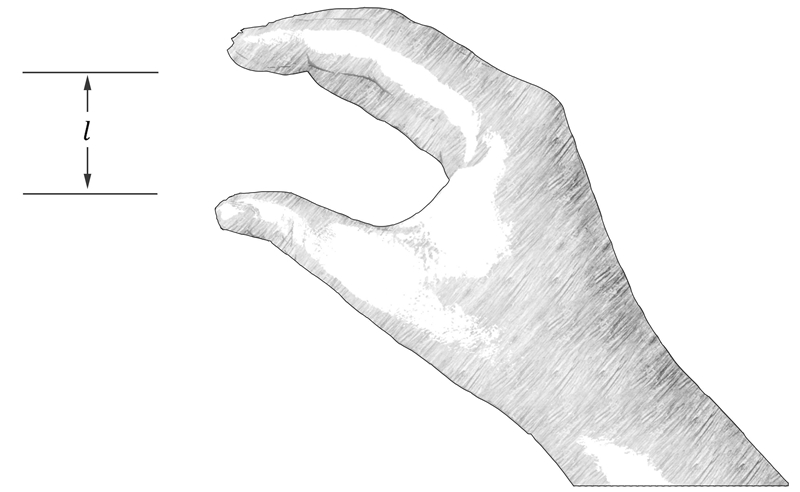}
    \caption{Finger Distance gesture ($l$ denotes the Euclidean distance between the fingertips of the index and the thumb of a tracked hand).}
    \label{fig:finger_distance}
\end{figure}
The Finger Distance gesture is a method that controls locomotion speed via the Euclidean distance between the fingertips of a user’s thumb and index. This was inspired by previous work by \citet{huang_design_2019} who used the distance between a user’s fingertip of the index and the hand centre to control walking speed. We decided to use the Euclidean distance between the tips of thumb and index fingers to calculate walking speed. Compared to the previous work, our method does not need to capture an initial resting hand gesture for calibration before locomotion starts. Thus, the present method is easier to set up and to use. The locomotion speed of our method is calculated by the following equations:
\begin{equation}
s_{walk} = \frac{l-d}{r-d}(s_{max}-s_{min})+s_{min} 
\end{equation}
\begin{equation}
l = \left \|F_{thumb} - F_{index}\right \|_2
\end{equation}

\noindent where $s_{walk}$ is the calculated walking speed, $s_{max}$ and $s_{min}$ are pre-defined maximum and minimum walking speeds that a user can achieve during locomotion, $r$ ($r$ = 8 cm) a pre-defined reference Euclidean distance value between fingertips of the thumb and the index of a hand, $d$ ($d$ = 2.5 cm) the dead zone that ensures a value close to zero can be obtained when users snap their thumb and index fingers together, $F_{thumb}$ and $F_{index}$ the tracked 3-D positions of the thumb and the index fingers and $l$ ($l\in(d,r]$) the calculated Euclidean distance between these two fingers. The values of $r$ and $d$ were empirical values determined during our initial testing. We found such choice of values gave a relatively comfortable feeling to control locomotion speed using the gesture. Whenever $l$ is less than or equal to $d$, the calculated walking speed $s_{walk}$ is set to zero to stop locomotion. An illustration of the proposed gesture is given in Figure~\ref{fig:finger_distance}, in which the distance between the tips of the thumb and index is depicted. This simple equation set enabled accurate control of locomotion speed. Locomotion speed changes linearly in relation to the distance between the thumb and index fingertips. Filtering is not necessary to pre-process the data of the fingertips as we use the immediate tracking data available from the Leap Motion sensor to calculate the distance between tips of thumb and index fingers. 

\begin{figure*}[!ht]
    \centering
    \subfigure[Stop (0 km/h)]{
        \includegraphics[width=0.2\textwidth]{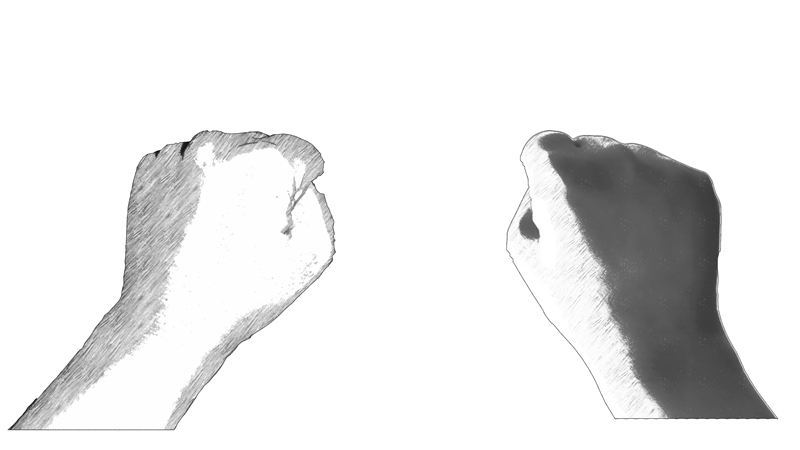}
    }
    \subfigure[1 km/h]{
        \includegraphics[width=0.2\textwidth]{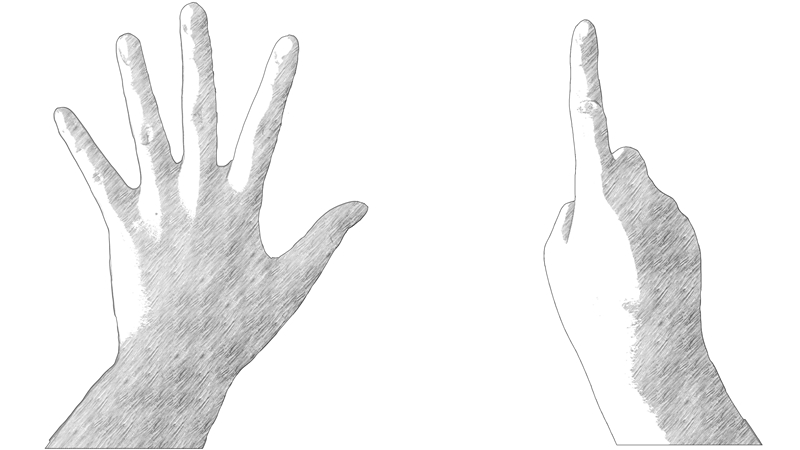}
    }
    \subfigure[2 km/h]{
        \includegraphics[width=0.2\textwidth]{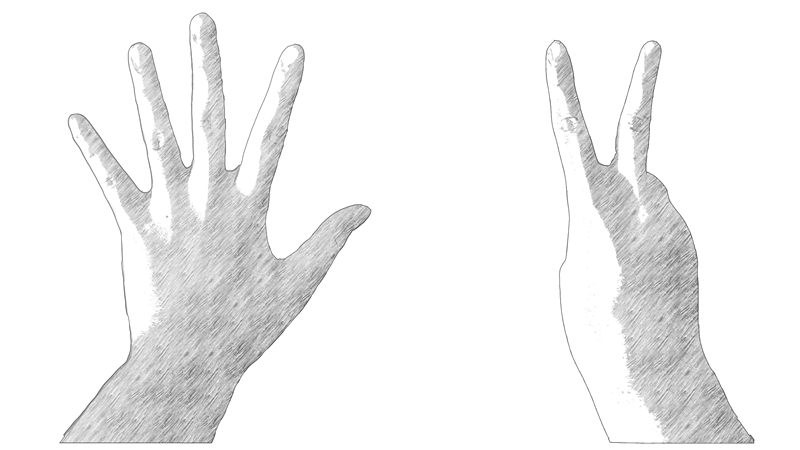}
    }
    \subfigure[3 km/h]{
        \includegraphics[width=0.2\textwidth]{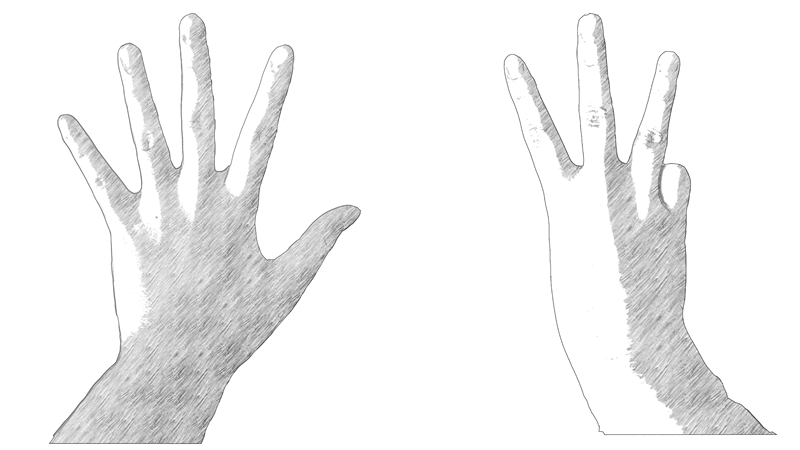}
    }
    \subfigure[4 km/h]{
        \includegraphics[width=0.2\textwidth]{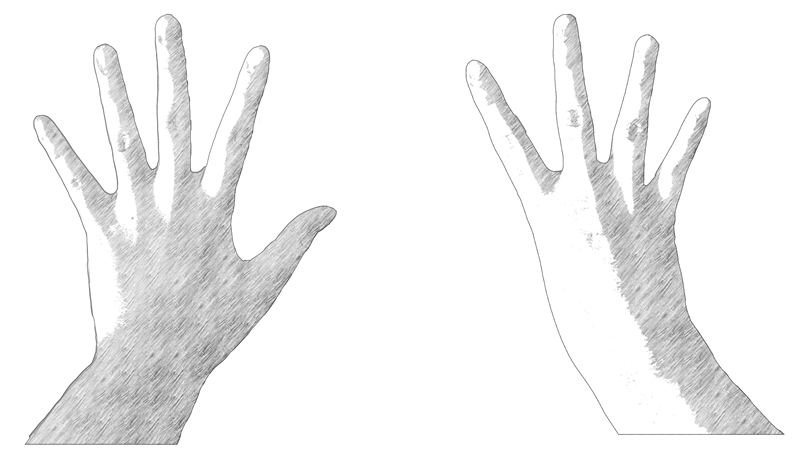}
    }
    \subfigure[5 km/h]{
        \includegraphics[width=0.2\textwidth]{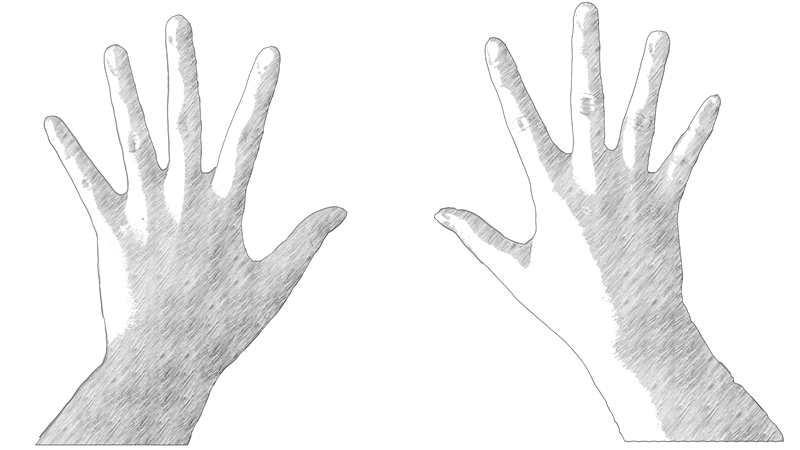}
    }
    \caption{Finger Number gesture. Sub-figures (a) - (f) illustrate six different gestures that correspond to walking speeds from 0 km/h to 5 km/h.}
    \label{fig:finger_number}
\end{figure*}

\subsection{Finger Number Gesture}
The Finger Number gesture is a set of gestures that controls the locomotion speed using the number of extended fingers. This idea was originally proposed by \citet{cardoso_comparison_2016} but details of the algorithms were not described in their paper. To design and implement an algorithm that can be used to recognize the number of extended fingers. We resorted to the robust feature descriptors by \citet{marin_hand_2016}, which normalize the distance between fingertips and the hand centre with respect to different hand orientations and varied distance to the Leap Motion sensor:

\begin{equation}
P_i^x=(F_i-C)\cdot(N\times H)
\end{equation}
\begin{equation}
P_i^y=(F_i-C)\cdot H
\end{equation}
\begin{equation}
P_i^z=(F_i-C)\cdot N
\end{equation}

\noindent where $P_i^x$, $P_i^y$ and $P_i^z$ are the extracted feature of a tracked 3-D fingertip position $F_i$ ($i$ is the index of a tracked finger), $C$ the tracked hand centre, $N$ the normal perpendicular to the palm of a tracked hand and $H$ the pointing direction of fingertips given by the Leap Motion sensor. $\cdot$ and $\times$ denote dot product and cross product, respectively. The set of descriptors forms a column vector of fifteen elements of a single tracked hand. Since the control of the start and the stop of locomotion involves both hands (\textit{i.e.} extend fingers to start walking and make fists with both hands to stop), we use the set of descriptors to extract features of both hands and stack the column vectors of features of both hands to form a vector $P$ of thirty elements. This compact feature representation enabled us to train a multi-class Support Vector Machine (SVM) using the LibSVM library \citep{chang_libsvm:_2011}, which makes it possible to recognize six types of hand gestures for controlling start and stop (\textit{i.e.} by making fists with both hands) and control locomotion speeds with five different levels (from 1 km/h to 5 km/h based on the number of extended fingers as shown in Figure~\ref{fig:finger_number}). Our definition of the hand gestures with respect to walking speed is different compared to that of Cardoso’s method. In his method, starting and stopping locomotion is controlled by extending both hands and closing both hands. In our implementation, extending an arbitrary number of fingers of the right hand (with the left hand fully extended) will start walking with the speed set by the number of extended fingers of the right hand (also see Figure~\ref{fig:finger_number}). To train a multi-class SVM for gesture recognition, we collected hand gesture data from twelve volunteers (age: 18-21, 6 males and 6 females) using a custom software application developed in Python 3.7 that enabled data collection and labelling. For each participant, we collected four sessions of data. During each data collection session, each gesture was collected for 5 s. We used the first two sessions of hand gesture data of all participants to train a multi-class SVM with the Radial Basis Function (RBF) kernel. We then used the remaining two sessions of the collected data to test the classifier. Overall, the average accuracy of classification is 99.94\%, which made it readily available for recognizing the Finger Number gesture for virtual locomotion using the $svm\_predict()$ function in the LibSVM library:

\begin{equation}
s_{walk} = svm\_predict(model, P)
\end{equation}

\noindent where $s_{walk}$ is the calculated walking speed, $model$ the trained multi-class SVM model and $P$ the stacked vector of features extracted from both hands using Marin et al.'s feature descriptor \citep{marin_hand_2016}. 

\begin{figure*}[!ht]
    \centering
        \includegraphics[width=1\textwidth]{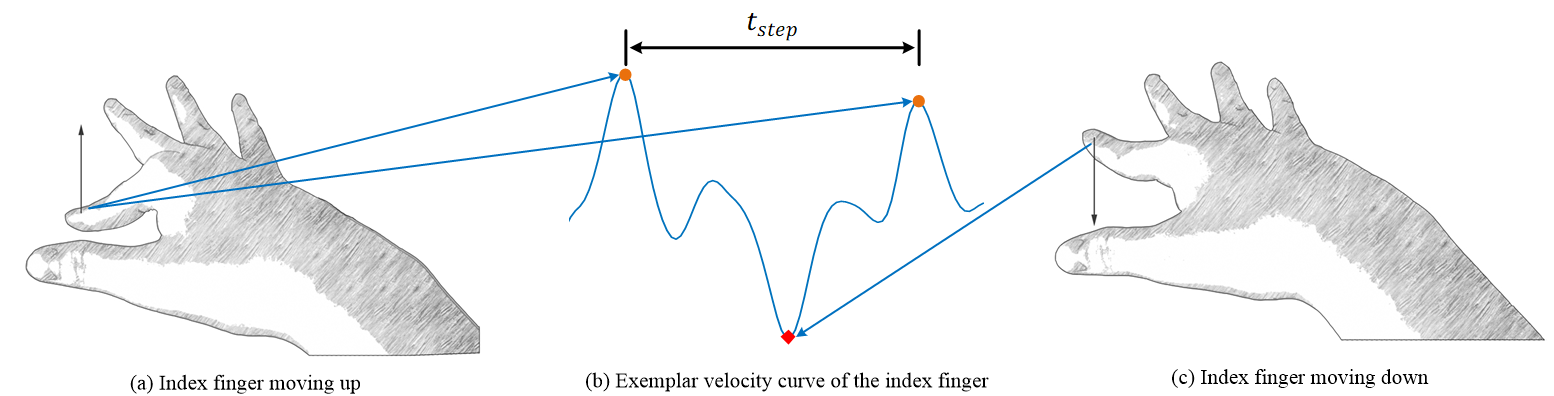}
    \caption{Finger Tapping gesture. Black arrows in sub-figures (a) and (c) illustrate the upward and downward motions of the index finger of a tracked hand during locomotion. Sub-figure (b) shows an exemplar velocity curve of the index finger in the vertical direction ($y$-axis). Orange round shapes represent the instant when the index finger reaches maximum velocity when moving up and the red diamond shape represents the instant when the index finger reaches minimum velocity when moving down. $t_{step}$ is the detected time interval between two detected peaks.}
    \label{fig:finger_tapping}
\end{figure*}

\subsection{Finger Tapping Gesture}
The Finger Tapping gesture is a method that controls locomotion speed via tapping (up and down) motions of the index finger of a tracked hand (see Figure~\ref{fig:finger_tapping} for the illustration of the proposed gesture). Our initial idea was to use two fingers (index and middle fingers) to simulate the walking motion of two legs during actual locomotion, similar to the Finger Walking in Place (FWIP) method  by \citet{kim_finger_2008,kim_effects_2010} and use the Leap Motion sensor to track finger motions instead of a multi-touch pad. Our initial design showed that using two fingers to simulate walking motion and control locomotion speed was difficult to achieve as moving one finger (index finger) also affected the tracked fingertip position of the other (middle finger). Thus, we decided to use the tapping motion of a single finger (index finger) to control walking speed. The tapping motion of the tip of the index finger can be viewed as a 3-D signal in the time domain. As the gesture requires a user to tap their fingers up and down, with their hands positioned above the Leap Motion sensor with a certain distance, we only needed the $y$-component of the tracked fingertip velocity to detect the time interval between two consecutive peaks and use that time interval to control walking speed. \citet{zhao_learning_2016} proposed gait analysis algorithms for real-time virtual locomotion using a treadmill with a large-scale stereo projected display and also adapted the algorithm for offline gait analysis to study the role of stereoscopic viewing in virtual locomotion \citep{zhao_role_2020}. This method works by first filtering foot motion signals (speed signals for real-time application or position signals for offline gait analysis) with low-pass Butterworth filters. Then, an empirical threshold is used as the starting point for finding the initial swing and terminal swing of a step through gradient descent. The searching stops whenever the gradient ascends (indicating a different gait cycle is detected) or the minimum threshold is reached. The index of the maximum value between the initial swing and the terminal swing is considered as the mid-swing. We considered the problem of detecting the peaks of the index fingertip speed signal as finding the mid-swings of steps. In the present implementation, a 2nd order low-pass Butterworth filter with a cut-off frequency of 5 Hz was applied to the $y$-component of the speed signal (buffered for 1 s using a queue) of the index finger. By adapting the algorithm, we were able to detect the peaks between consecutive steps and calculate the time interval between consecutive peaks. The equation to obtain the calculated walking speed is given by:

\begin{equation}
s_{walk}=(1-\frac{t_{step}-t_{min}}{t_{max}-t_{min}})(s_{max}-s_{min})+s_{min}
\end{equation}

\noindent where $s_{walk}$ is the calculated walking speed, $t_{step}$ ($t_{step}\in(t_
{min},t_{max}]$) the time interval between two detected peaks, $t_{max}$ ($t_{max}$ = 0.95 s) and $t_{min}$ ($t_{min}$ = 0.3 s) the pre-defined maximum and minimum time intervals and $s_{max}$ and $s_{min}$ the pre-defined maximum and minimum walking speeds that one can achieve using the method. A larger $t_{step}$ due to slow tapping motion results in a slower walking speed $s_{walk}$, which makes it possible for speed control. This equation is similar to the approach by \citet{tregillus_vr-step:_2016}, in which a WIP method was proposed based on the acceleration signals from a mobile phone to detect steps and calculate walking speed based on the time intervals between steps. Additionally, whenever the time interval $t_{step}$ is larger than 0.95 s, the calculated walking speed  $s_{walk}$ is set to $s_{min}$ and whenever $t_{step}$ is less than or equal to 0.3 s, the calculated walking speed is set to $s_{max}$.

Our current technique is based on tracking only the index finger due to the technical limitation of the Leap Motion sensor. If there are new hand-tracking sensors available in the future, tracking both index and middle fingers while moving them alternately may be stable. Thus, simulating bipedal walking using index and middle fingers would be feasible. On the other hand, there are high-precision marker-based systems, such as the OptiTrack, that potentially enable this type of tracking. Adapting this Finger Tapping algorithm to work for two fingers is trivial. Both two index and middle fingers can adapt locomotion speed once the parameter $t_{step}$ is detected for either finger. This will make the Finger Tapping technique close to the metaphor of the Finger Walking in Place (FWIP) technique.

\subsection{Direction Control}
The three hand gesture interfaces presented in previous sections required users to use their right hands to control walking speed. To enable direction control during locomotion, our first thought was to use the same hand (right hand) to control both walking speed and walking direction. However, early testing showed that this was not possible as moving fingers to make a gesture with the right hand also changed the tracked hand pointing direction $H$ given by the Leap Motion sensor. Thus, we decided to let users control their walking direction using their left hands. The equations for calculating the steering angle are given by:

\begin{equation}
d=V_n\cdot(H_{init}\times H_{current})
\end{equation}
\begin{equation}
\theta=sign(d)cos^{-1}\frac{H_{init}\cdot H_{current}}{\left\|H_{init}\right\|\left\|H_{current}\right\|}
\end{equation}

where the $sign(x)$ function is defined as:
\begin{equation*}
sign(x) = 
\begin{cases}
&\text{-1,  } x<0 \\ 
&\text{ 1,  } x\ge0 
\end{cases}
\end{equation*}

\begin{figure}[t]
    \centering
        \includegraphics[width=0.35\textwidth]{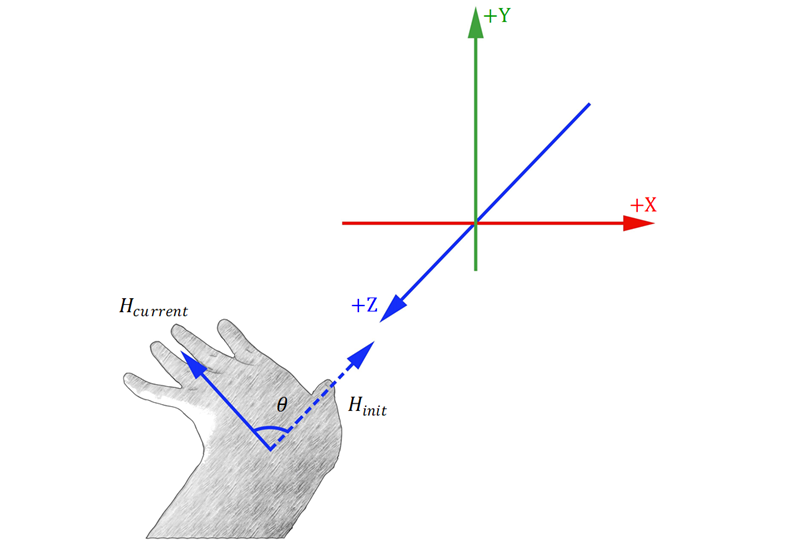}
    \caption{Direction control using the left hand, illustrated in the Leap Motion sensor's coordinate system ($\theta$ denotes the steering angle).}
    \label{fig:direction_control}
\end{figure}

\noindent Parameter $d$ (with a $sign(x)$ function) controls whether a user is turning left or turning right, $V_n$ the vector that is perpendicular to the $x$-$z$ plane ($V_n=(0,1,0)$), $H_{init}$ the initial vector that points to a user’s direction of travel ($H_{init}=(0,0,-1)$), $H_{current}$ the current finger pointing direction tracked by the Leap Motion sensor and $\theta$ the steering angle that a user intends to achieve. $H_{current}$ tracked by the Leap Motion sensor is slightly noisy, which made straight-line walking and turning both unstable. To ensure smoothness in linear walking and turning, we used the function $SmoothDamp()$ from the Unity's library to damp turning motions. An illustration of the direction control gesture is in Figure~\ref{fig:direction_control}. Users extend their fingers and rotate their palms around the $y$-axis to control their travel direction.

\begin{figure}[t]
    \centering
        \includegraphics[width=0.35\textwidth]{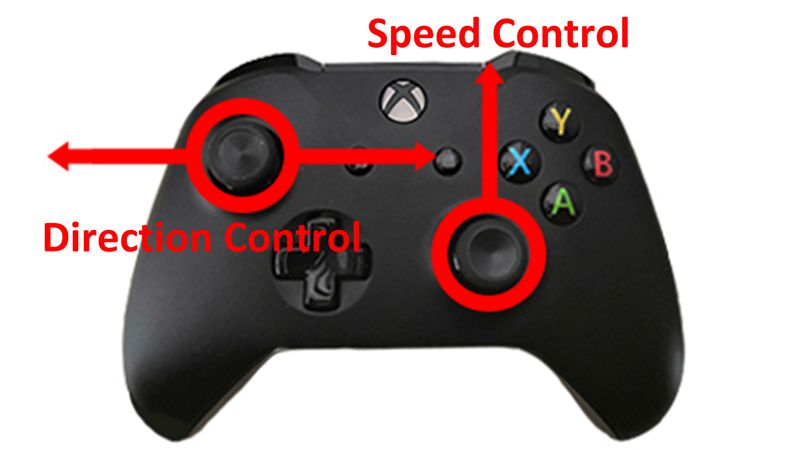}
    \caption{Gamepad interface based on the Xbox One controller. Pushing the left joystick left or right controls walking direction and pushing the right joystick forward controls walking speed.}
    \label{fig:gamepad}
\end{figure}

\subsection{Gamepad Interface}
The gamepad interface was implemented as a control interface to be compared to the three hand gesture interfaces. In our implementation, pushing the left joystick left and right controls steering and pushing the right joystick upward controls a user’s walking speed (see Figure~\ref{fig:gamepad} for illustration). Moving backward was not allowed for experiments so this function was disabled. As in previous gesture interfaces, participants used their left hands for direction control and right hands for speed control.

\begin{table}%
\centering
\caption{The User Interface Questionnaire}
\label{tab:user_interface}
\begin{tabular}{ p{8cm}}
\hline
The interface is easy to learn.\\
\hline
The interface is easy to use.\\
\hline
The interface is natural and intuitive to use.\\
\hline
The interface helps make the task fun.\\
\hline
Using the interface is tiring.\\
\hline
The interface helps me respond quickly.\\
\hline
The interface helps me make accurate responses.\\
\hline
\end{tabular}\
\end{table}

\begin{figure}[t]
\centering
\includegraphics[width=0.45\textwidth]{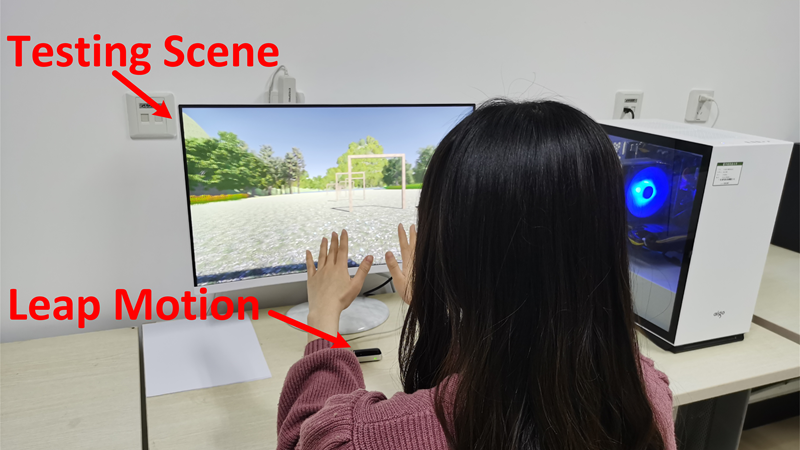}
\caption{Experimental setup. The participant sat in front of the 27-inch curved monitor with hands positioned above the Leap Motion sensor. The viewing distance was approximately 60 cm. The virtual scene displayed on the monitor captured in this figure was a snapshot of Experiment 2.}
\label{fig:setup}
\end{figure}

\subsection{User Interface Questionnaire} \label{sec:ui}
To subjectively evaluate four interfaces, we adapted the user interface questionnaire from \citet{nabiyouni_comparing_2015} and \citet{zhao_comparing_2019} and used the questionnaire in the present study. The questionnaire included statements, such as “The interface is easy to use” (see Table~\ref{tab:user_interface} for the complete list), for users to judge. For both experiments, after finishing using an interface for a block of the experiments, participants were asked to rate each factor presented in the questionnaire using a seven-point Likert scale (from strongly disagree to strongly agree) to indicate their preference. This user interface questionnaire allowed us to subjectively evaluate a given interface using factors, including ease-to-learn, ease-to-use, natural-to-use, fun, tiredness, responsiveness and subjective accuracy. These are typical factors for consideration when designing or evaluating new interfaces.

\subsection{Hardware and Software of the VR system}
The software application that included virtual scene presentation, gesture recognition, virtual locomotion tasks and data recording was developed using the Unity 2020.1. The software application was hosted on a Windows 10 desktop computer with an Intel i5-10500 CPU, 16 GB memory, an Nvidia Geforce 1660s graphics card with 6 GB graphics memory. Hand movements of participants were tracked by the Leap Motion sensor using the Orion SDK 4.0.0. The display was an AOC 27-inch curved monitor for presenting the virtual scenes of the tasks. The avatar that represented a participant in the virtual environment and for collision detection was implemented using a 3-D capsule geometry, with the virtual camera placed near the top of the capsule. When users made hand gestures, the walking speed and direction of the capsule were changed. The experimental setup for both experiments is shown in Figure~\ref{fig:setup}.

\section{Experiments}
\subsection{Experiment 1: Target Pursuit}
\subsubsection{Introduction}
The purpose of Experiment 1 was to assess the performance and user preference of four locomotion interfaces on speed control.

\subsubsection{Participants}
Sixteen undergraduate students (age: 18-25, eleven males and five females) were invited as volunteers for this experiment. All had normal or corrected-to-normal vision. An informed consent was signed before the experiment. Participants were naïve to the purpose of the experiment.

\begin{figure}[t]
\centering
\includegraphics[width=0.45\textwidth]{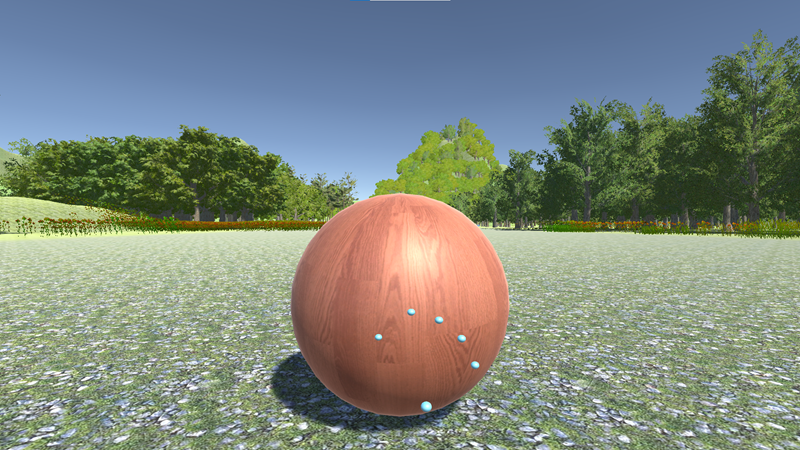}
\caption{Exemplar snapshot of Experiment 1. Light blue coloured spheres represent a user's tracked fingers and the hand centre. The wooden ball captured in this snapshot was red as the participant was close to the ball.}
\label{fig:ball}
\end{figure}

\subsubsection{Design}
We adapted the experiment from \citet{zhao_learning_2016} to assess the performance of speed control of four interfaces. We designed a virtual scene (shown in Figure~\ref{fig:ball}) with foliage and trees in the Unity 2020.1 using resources (the Free SpeedTrees Package and the Grass Flowers Pack Free) from the Unity Asset Store. A wooden ball was placed before the start of an experimental trial in the scene 3 m in front of the avatar that represented a participant. After a trial started, the rolling ball changed its forward running speed based on a set of randomly generated speed key frames (with values 2 km/h, 3 km/h or 4 km/h), which took effect every 10 s. In total, there were seventeen speed key frames for each trial. The goal of the experiment for participants was to pursue the rolling ball while maintaining the initial 3 m distance between the avatar and the rolling ball using each locomotion interface. Direction control was disabled for this task. A complete trial lasted 3 min and an experiment trial stopped after 3 min. To make the distance judgement easier for participants, we manipulated the color of the rolling wooden ball such that the color would become red if the avatar was getting close or blue if getting far away. The acceleration of the avatar was set to 0.5 ${\rm m/s^2}$ for all four interfaces. The maximum locomotion speed and the minimum locomotion speed were set to 5 km/h and 0 km/h, respectively, The linear acceleration of the wooden ball was set to 0.3 ${\rm m/s^2}$. Setting the maximum walking speed to 5 km/h and the acceleration to 0.5 ${\rm m/s^2}$ was reasonable as \citet{teknomo_microscopic_2002} reported that the average walking speed of pedestrian flow is 1.38 m/s $\pm$ 0.37 m/s (4.97 km/h $\pm$ 1.33 km/h, mean $\pm$ std) and the average acceleration is 0.68 ${\rm m/s^2}$. Light blue coloured spheres (as shown in Figure~\ref{fig:ball}) were added and rendered on top of all other geometries in the scene to visualise a user's tracked fingertips and the hand centre. This decision was made after our initial testing. We found it necessary to provide users with tracking information of their hands so that they knew their hands were securely tracked. To control for order effects, we counter-balanced the order of access to four different interfaces using the balanced Latin square design. 

\subsubsection{Metrics} \label{sec:exp1_metrics}
Based on the metrics used by \citet{zhao_learning_2016}, we proposed five metrics to study the performance of the interfaces. These included:
\begin{itemize}
\item \textbf{Average position difference \bm{$d_{avg}$}}:\\
The average distance between the avatar and the rolling ball through the entire course. Ideally, the value should be 3 m as participants were required to maintain this distance between the avatar and the rolling ball through an entire course.
\item \textbf{Standard deviation of position difference \bm{$d_{std}$}}:\\
The standard deviation of the distance between the avatar and the rolling ball through the entire course. This parameter reflects the interval, in which the avatar oscillates while maintaining the 3 m distance.
\item \textbf{Average speed difference \bm{$s_{avg}$}}:\\
The speed difference $s_{avg}$ between the avatar and the rolling ball through the entire course. Ideally, the value should be zero assume that participants are able to perfectly maintain the 3 m distance through the entire course.
\item \textbf{Standard deviation of speed difference \bm{$s_{std}$}}:\\
The standard deviation of the speed difference between the avatar and the rolling ball through the entire course. This parameter reflects the interval in which the avatar varies their locomotion speed while maintaining the 3 m distance.
\item \textbf{Speed difference at key frames \bm{$s_{inst}$}}:\\
The average speed difference between the avatar and the wooden ball from the instant the speed change of the wooden ball occurs to 0.1 s after this speed change has occurred. This parameter reflects the transient response of locomotion interfaces when being used for tracking speed changes.
\end{itemize}

\subsubsection{Procedure}
Both participants and researchers wore masks during experiments. Contact surfaces were cleaned using alcohol wipes before and after each experiment. During an experimental session, a participant was first introduced to the task and asked to sign an informed consent. Then, participants sat in front of the curved monitor with their hands positioned above the Leap Motion sensor. A researcher sat beside the participant and operated the experimental software to initiate new experimental trials. Participants were asked to perform one trial as practice to get familiar with the interface. After that, participants were asked to perform three experimental trials using the same interface, with their locomotion data recorded during the trials. The order of access to interfaces was determined using the balanced Latin square design. When participants completed a block (consisting of a practice trial and three experimental trials) for a given interface, they were asked to rate the interface using the user interface questionnaire introduced in Section~\ref{sec:ui}. They were subsequently introduced to another interface, asked to complete a practice trial and three experimental trials and fill in another user interface questionnaire. We ran the same procedure until a participant used all four interfaces to complete the experiment. On average, an entire experimental session took an hour to complete, taking into account the time durations to complete 16 trials (each lasted 3 min) and fill four questionnaires in total.

\begin{figure*}[!ht]
\centering
\includegraphics[width=0.85\textwidth]{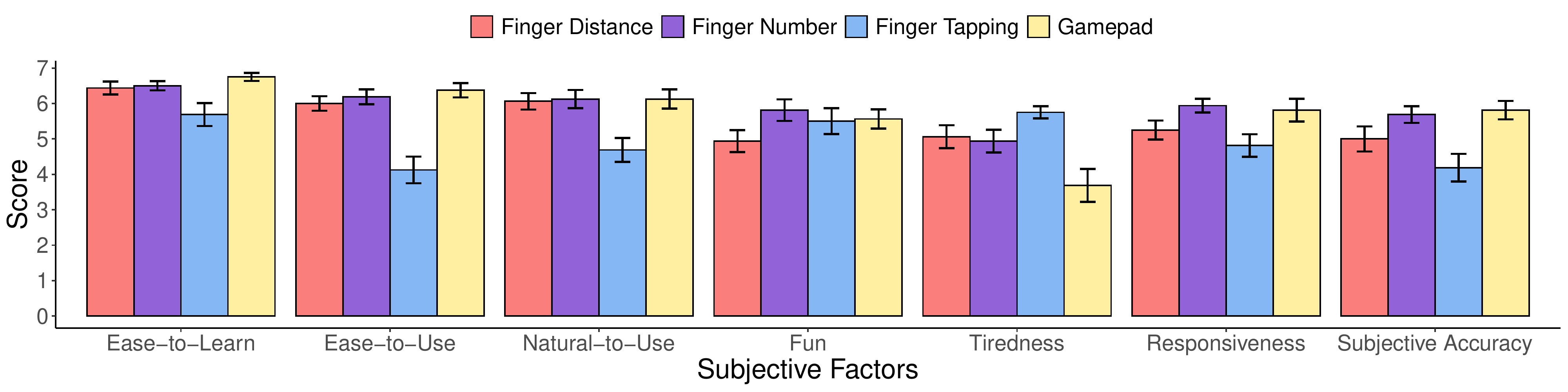}
\caption{Subjective factors of Experiment 1 (bars denote mean values and error bars denote the standard error of the mean).}
\label{fig:ball_errorbar}
\end{figure*}

\begin{figure*}[!ht]
\centering
\includegraphics[width=0.85\textwidth]{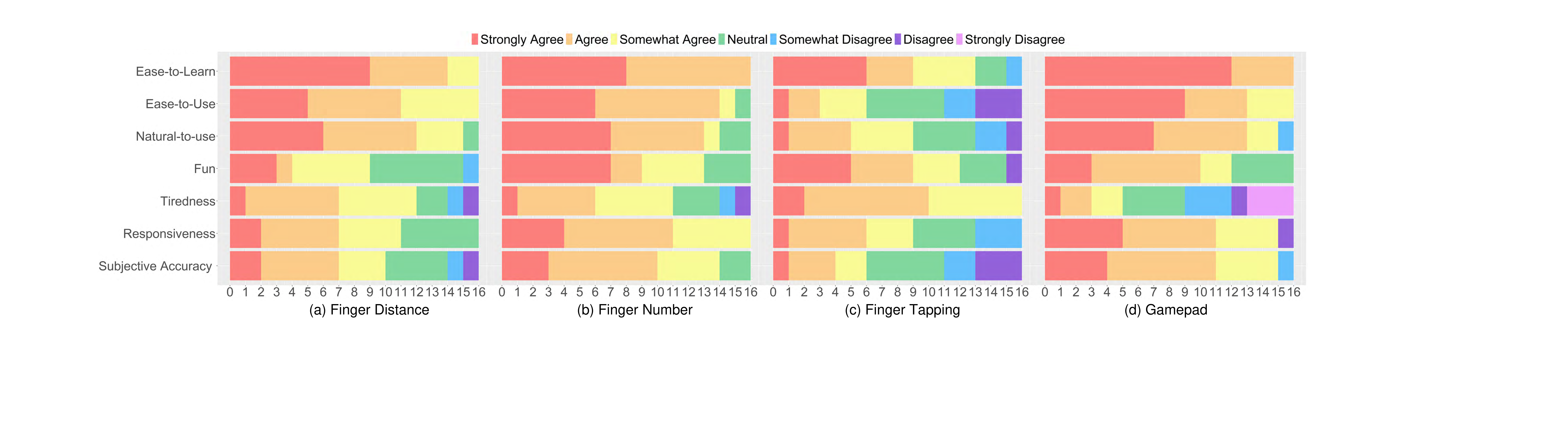}
\caption{Detailed responses from participants in Experiment 1.}
\label{fig:ball_horizontal}
\end{figure*}

\begin{figure*}[!ht]
\centering
\includegraphics[width=0.85\textwidth]{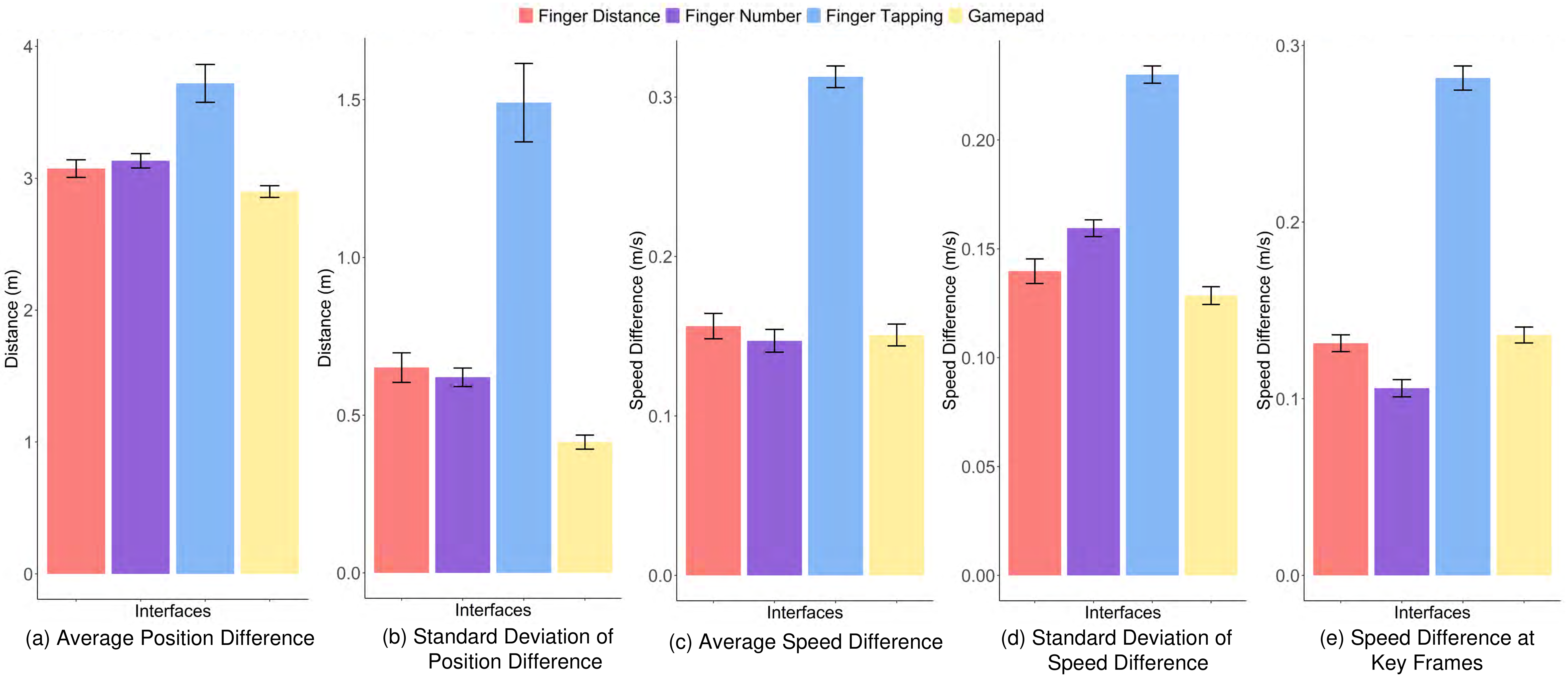}
\caption{Objective factors of Experiment 1 (bar plot convention is as in Figure~\ref{fig:ball_errorbar}).}
\label{fig:objective_ball}
\end{figure*}

\begin{table*}[!ht]%
\centering
\caption{Results of Statistical Analyses on Subjective Factors of Experiment 1}
\label{tab:ball_subjective}
\includegraphics[width=0.8\textwidth]{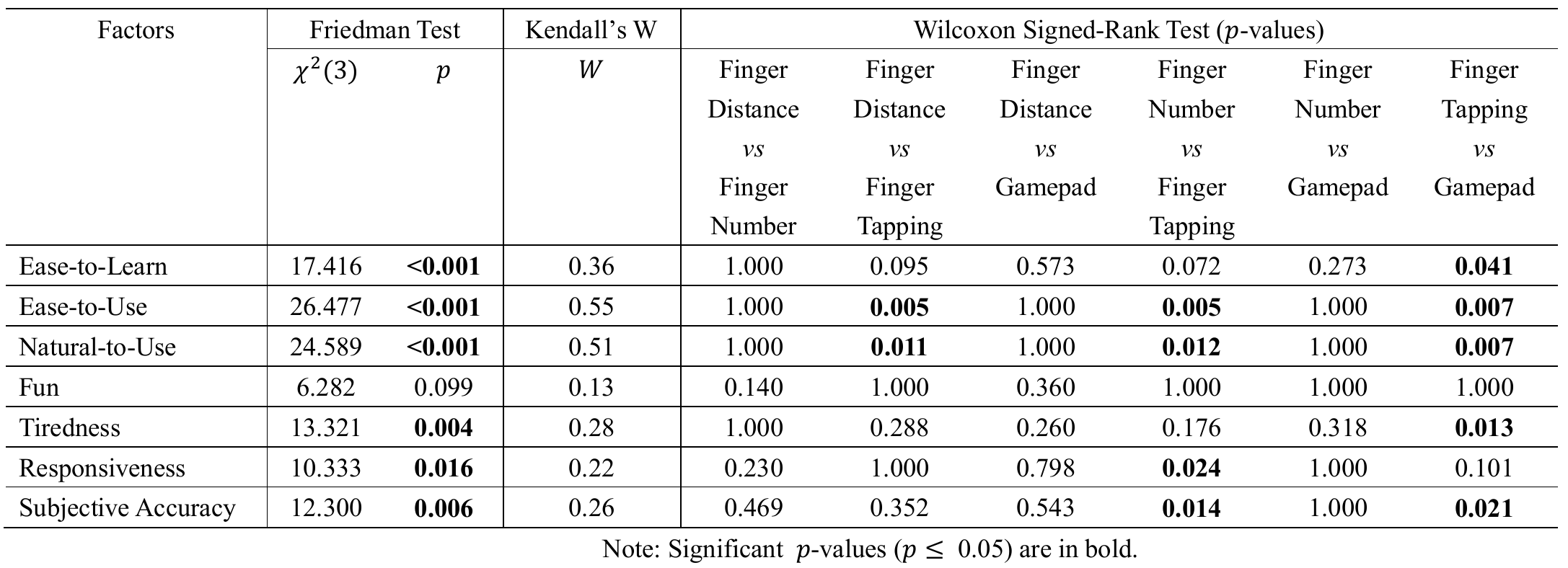}
\end{table*}

\begin{table*}[!ht]%
\centering
\caption{Results of Statistical Analyses on Objective Factors of Experiment 1}
\label{tab:ball_objective}
\includegraphics[width=0.6\textwidth]{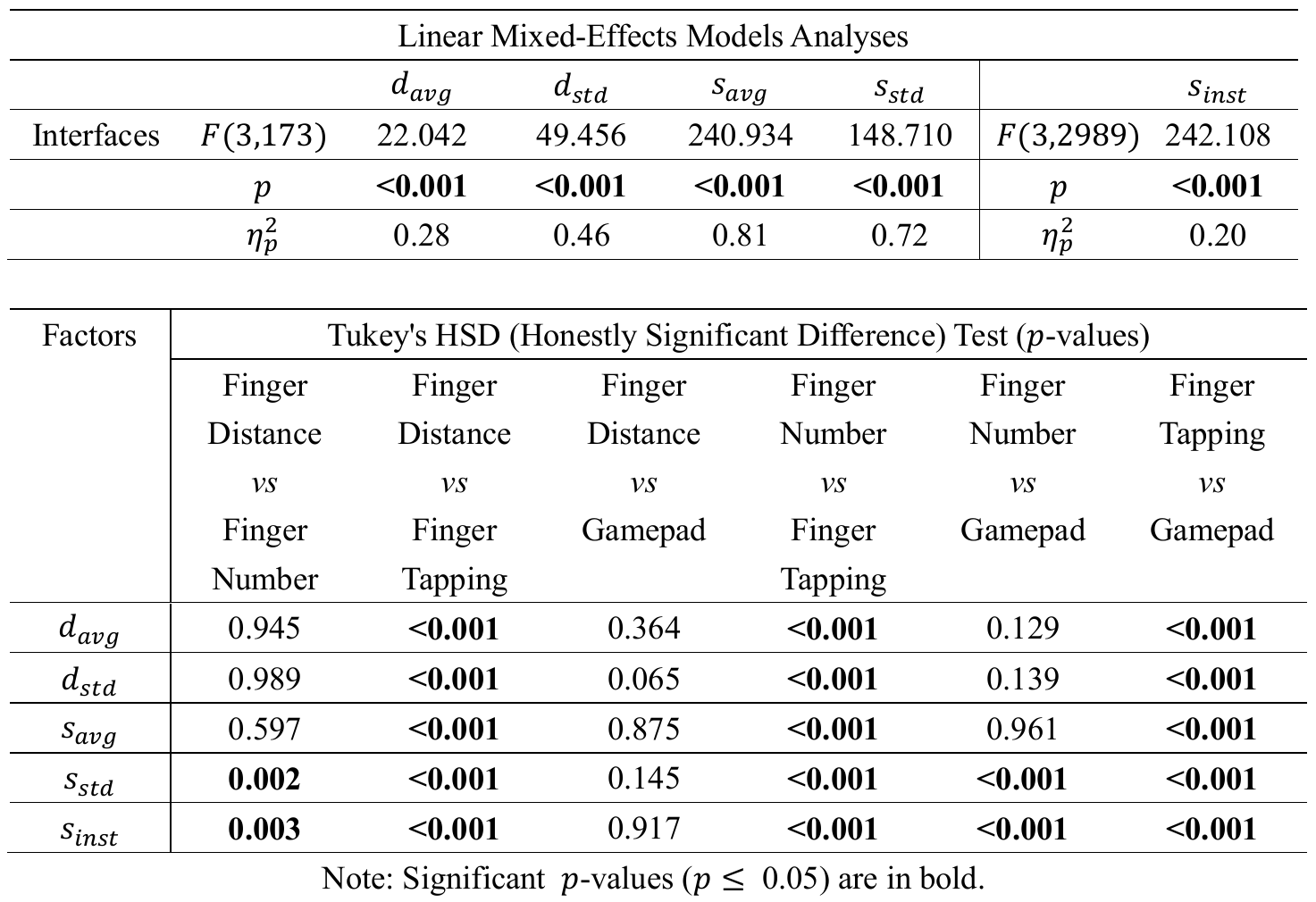}
\end{table*}

\subsubsection{Results}
We performed data analyses using R 4.0.3. The independent factor of the statistical analyses was the interfaces (Finger Distance, Finger Number, Finger Tapping and Gamepad) and was treated as a fixed effect. The dependent factors for subjective analyses were the factors given by the user interface questionnaire. The dependent factors for objective analyses were the parameters defined in Section~\ref{sec:exp1_metrics}. Participants were treated as a random effect. Subjective data given by participants using the user interface questionnaire were analysed using the Friedman test, followed by the Wilcoxon signed-rank test with the Bonferroni correction as the post-hoc analyses. Effect sizes were estimated using the Kendall's Coefficient of Concordance (W). Figure~\ref{fig:ball_errorbar} shows the mean values and the standard error of the mean and Figure~\ref{fig:ball_horizontal} shows the distribution of the responses from participants. Objective parameters based on the metrics defined in Section~\ref{sec:exp1_metrics} were extracted from the collected experimental data and were analysed using the linear mixed-effects models analyses (Package NLME in R). The post-hoc analyses for objective data were performed using the Tukey's HSD (Honestly Significance Difference) test. Effect sizes were reported using partial eta squared ${\eta}_{p}^{2}$ (estimated from repeated-measures ANOVA analyses of the same form as the linear mixed-effects models analyses). No significant learning effect resulting from the number of trials was found on all objective factors. The error contribution of the number of trials was minimal so it was not included in the analyses. Figure~\ref{fig:objective_ball} shows the mean and the standard error of the mean of all objective parameters. Results of the statistical analyses on both subjective data and objective data are in Table~\ref{tab:ball_subjective} and Table~\ref{tab:ball_objective}, respectively.

Analyses on subjective parameters using the Friedman test showed that interfaces had significant effects on factors including ease-to-learn ($p<0.001$), ease-to-use ($p<0.001$), natural-to-use ($p<0.001$), tiredness ($p=0.004$), responsiveness ($p=0.016$) and subjective accuracy ($p=0.006$) except fun ($p=0.099$). Pairwise comparisons were performed using the Wilcoxon signed-rank test with the Bonferroni correction on subjective factors. Results showed that Gamepad was significantly easier to learn compared to Finger Tapping. Although Finger Tapping had a lower mean score on ease-to-learn, there was no statistical significance compared to Finger Distance and Finger Number. The result was reasonable as the Gamepad interface is a traditional device for navigation in games so participants found it easier to learn compared to other interfaces while hand gesture interfaces were all relatively new to them. Finger Distance, Finger Number and Gamepad were found to be significantly easier to use compared to Finger Tapping and there was no significant effect among Finger Distance, Finger Number and Gamepad on ease-to-use. A primary reason was that the control of walking speed through Finger Tapping could only be made after two peaks on the velocity curve of the index finger were detected so there was a latency in adjusting walking speed compared to other interfaces. We also found that Finger Distance, Finger Number and Gamepad were significantly natural to use compared to Finger Tapping and there was no significant effect among Finger Distance, Finger Number and Gamepad on natural-to-use. This showed that Finger Tapping was somewhat unusual to participants and participants needed to take practice to get familiar with the interface. Therefore, lower ratings were received for this factor. There was no statistical significance among these interfaces in terms of fun, but relatively high mean scores on this parameter indicated all four interfaces were of fun for virtual walking in VR. Finger Tapping was found to be significantly more tiring compared to Gamepad while there was no significant difference among Finger Distance, Finger Number and Finger Tapping, showing that repetitive motions from Finger Tapping caused fatigue. Finger number was found to be significantly more responsive compared to Finger Tapping and there was no statistical significance among Finger Number, Finger Distance and Gamepad. This was also due to the latency problem inherent in the current design of the Finger Tapping gesture. Finally, Finger Number and Gamepad were found to be significantly more subjectively accurate compared to Finger Tapping. There were no statistical significance between Finger Number and Gamepad, and no statistical significance between Finger Distance and Finger Tapping in terms of subjective accuracy, indicating that Finger Number and Gamepad were considered more accurate by participants. By examining Figure~\ref{fig:ball_horizontal}, we found that Gamepad had the most number of ratings of strongly agree excluding tiredness. Finger Number had a similar response in terms of the number of ratings of strongly agree and was slightly higher than that of Finger Distance. Finger Tapping received the lowest number of ratings of strongly agree. 

Analyses on objective factors using the linear mixed-effects models analyses showed that there were significant effects on all five parameters, including average distance $d_{avg}$ ($p<0.001$), standard deviation of distance $d_{std}$ ($p<0.001$), average speed difference $s_{avg}$ ($p<0.001$), standard deviation of speed difference $s_{std}$ ($p<0.001$) and speed difference at key frames $s_{inst}$ ($p<0.001$). Further analyses using Tukey's HSD test (with examination of Figure~\ref{fig:objective_ball}) showed that Finger Tapping had a larger error in terms of maintaining the 3 m distance between the avatar and the rolling wooden ball. The performance of Finger Distance, Finger Number and Gamepad were similar for this parameter. This showed that Finger Distance, Finger Number and Gamepad were more accurate in maintaining a relative distance to the rolling ball. Similar results were also found on parameters, including standard deviation of distance $d_{std}$ and average speed difference $s_{avg}$. Finger Tapping was found to have a larger oscillating interval $d_{std}$ in terms of the distance between the avatar and the rolling wooden ball. Finger Tapping was also found to have a larger error in terms of average speed difference $s_{avg}$ during the target pursuit task. In terms of standard deviation of speed difference $s_{std}$, Finger distance and Gamepad had similar performance and the values were significantly lower than that of the rest two interfaces. Finger Number was also found to have a significantly lower value for standard deviation of speed difference $s_{std}$ compared to Finger Tapping. In terms of speed difference at key frames $s_{inst}$, Finger Distance had similar performance compared to Gamepad. Finger Number was significantly better than other interfaces and Finger Tapping had significantly lower performance. This showed that Finger number has the best transient response in terms of tracking speed changes.

\subsubsection{Summary}
Experiment 1 showed that the Finger Number gesture and the Finger Distance gesture had similar user preference compared to the Gamepad interface. The Finger Tapping gesture received lower ratings for all factors except Fun. In terms of their performance on speed control, the Finger Distance gesture was close to the Gamepad method. The Finger Number gesture was slightly worse in terms of standard deviation of speed difference and speed difference at key frames  compared to the Finger Distance gesture and the Gamepad interface. The Finger Tapping gesture had largest errors in terms of maintaining the distance and speed.

\begin{figure}
\centering
\includegraphics[width=0.45\textwidth]{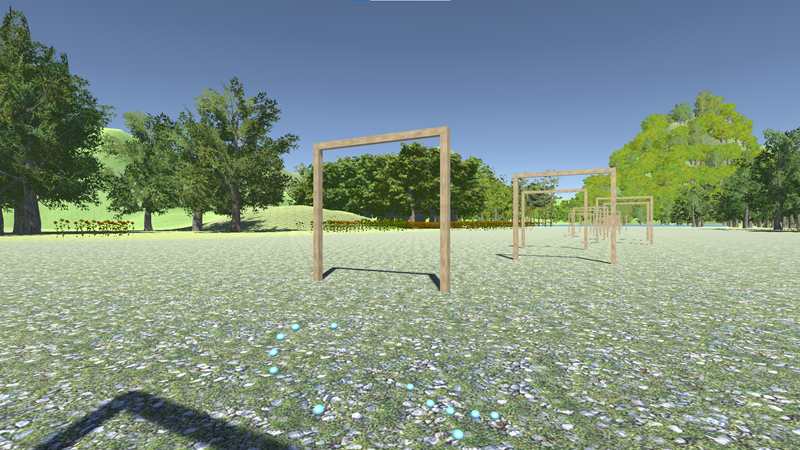}
\caption{Exemplar snapshot of Experiment 2. Wooden gates are the waypoints. Light blue coloured spheres represent a user's tracked fingers and hand centres of both hands.}
\label{fig:gate}
\end{figure}

\subsection{Experiment 2: Waypoints Navigation}
\subsubsection{Introduction}
The purpose of Experiment 2 was to test the performance and user preference of four interfaces in terms of waypoints navigation when direction control using the left hand was introduced. While Experiment 1 focused on speed control using hand gestures and the gamepad interface, Experiment 2 was a more general locomotion task that allowed participants to use their preferred locomotion speed with direction control to complete their task.

\subsubsection{Participants}
Sixteen undergraduate students (age: 18-24, eight males and eight females) were invited as volunteers for this experiment. None had participated in Experiment 1 and all had normal or corrected-to-normal vision. Informed consents were signed before experiments. Participants were naïve to the purpose of the experiment.

\subsubsection{Design}
We adapted this experiment from \citet{zhao_effects_2018} to evaluate the usability of interfaces for waypoints navigation. In this experiment, we used the same terrain of Experiment 1 but placed wood gates as the waypoints in the virtual environment (shown in Figure~\ref{fig:gate}). These wood gates had an inner dimension of 2 m (W) $\times$ 2 m (H) $\times$ 0.1 m (D). The backside of a gate to the front of its successor was fixed as 5 m in depth. In the lateral direction, these gates were randomly placed in an interval of $\pm$ 2 m before each trial of the experiment. In total, fifty waypoints were placed in the scene. The initial distance from the avatar to the front side of the first waypoint was 5 m. The distance between the backside of the last waypoint to the bounding box that triggered the end of the experiment was also 5 m. An experimental trial stopped whenever a participant reached the bounding box placed behind the last waypoint. The goal for participants of the experiment was to navigate through waypoints using four different interfaces while trying to achieve short task completion time and fast locomotion speed.  It was also necessary to avoid missing waypoints or colliding with waypoints. Participants had direction control and speed control using both hands when using each of the four interfaces to navigate through waypoints. In addition, participants were not allowed to backtrack a waypoint in case they missed. The acceleration of the avatar was also set to 0.5 ${\rm m/s^2}$, the maximum and minimum walking speeds were set to 5 km/h and 0 km/h, respectively, for all four interfaces. For this experiment, we also adopted the balanced Latin square design to control for order effects. 

\subsubsection{Metrics}\label{sec:exp2_metrics}
Based on the metrics given by \citet{zhao_effects_2018}, we proposed the following metrics to evaluate the interfaces for our study:
\begin{itemize}
\item \textbf{Task completion time \bm{$t_c$}}:\\
The duration from the start of the locomotion to the instant when participants reach the bounding box that triggers the end of the task.
\item \textbf{Average locomotion speed \bm{$s_l$}}:\\ 
The average locomotion speed computed by dividing the total length of the locomotion path by task completion time \bm{$t_c$}.
\item \textbf{Smoothness of the locomotion path \bm{$d_p$}}:\\
The mean value of the lateral distance ($x$-axis) of the actual locomotion path to the optimal locomotion path. The optimal locomotion path is defined as the shortest distance between the centres of two adjacent waypoints. 
\item \textbf{Number of successfully passed waypoints \bm{$n_w$}}:\\
The number of waypoints that participants successfully passed.
\item \textbf{Number of collisions with waypoints \bm{$n_c$}}:\\
The number of the collisions of the avatar with the frames of the waypoints. This parameter was obtained from collision detection of the Unity during gameplay and was recorded for analysis.
\end{itemize}

\subsubsection{Procedure}
The procedure was as in Experiment 1 except that the waypoints navigation task was given to the participants.

\begin{figure*}[!ht]
\centering
\includegraphics[width=0.85\textwidth]{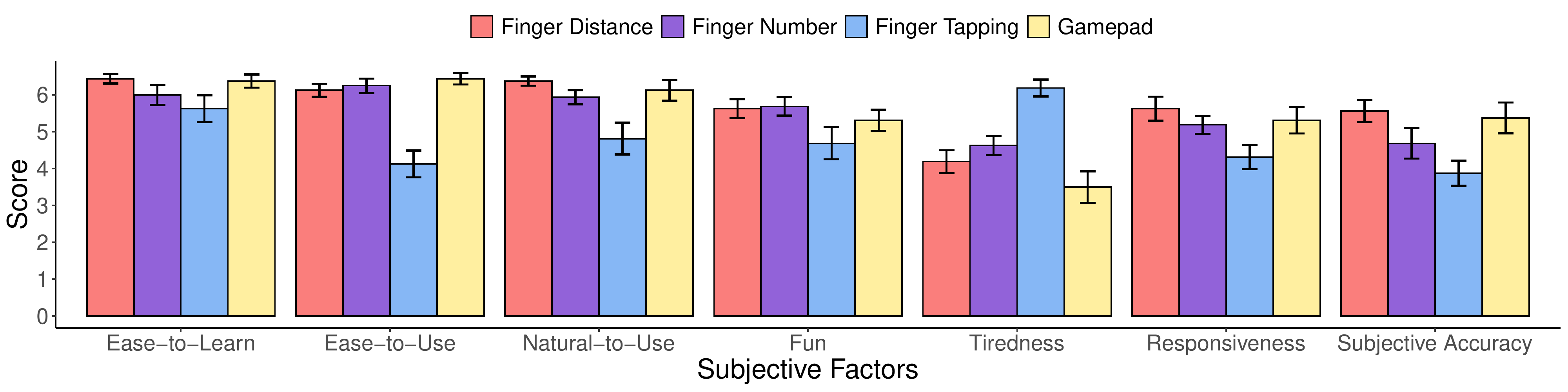}
\caption{Results of subjective factors of Experiment 2 (bar plot convention is as in Figure~\ref{fig:ball_errorbar}).}
\label{fig:door_errorbar}
\end{figure*}

\begin{figure*}[!ht]
\centering
\includegraphics[width=0.85\textwidth]{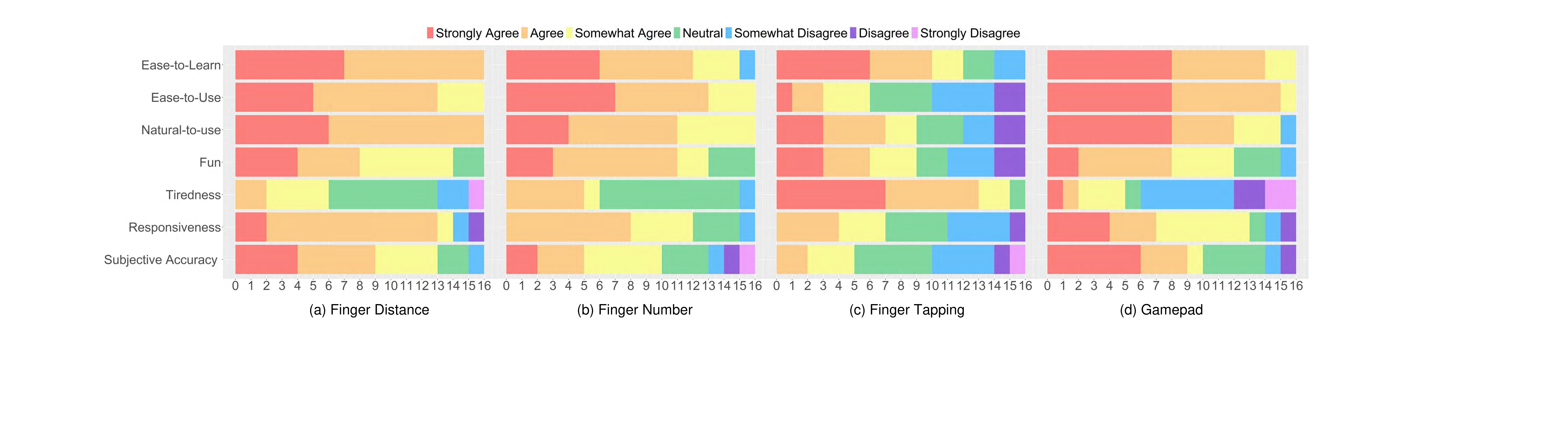}
\caption{Detailed responses from participants in Experiment 2.}
\label{fig:door_horizontal}
\end{figure*}

\begin{figure*}[!ht]
\centering
\includegraphics[width=0.85\textwidth]{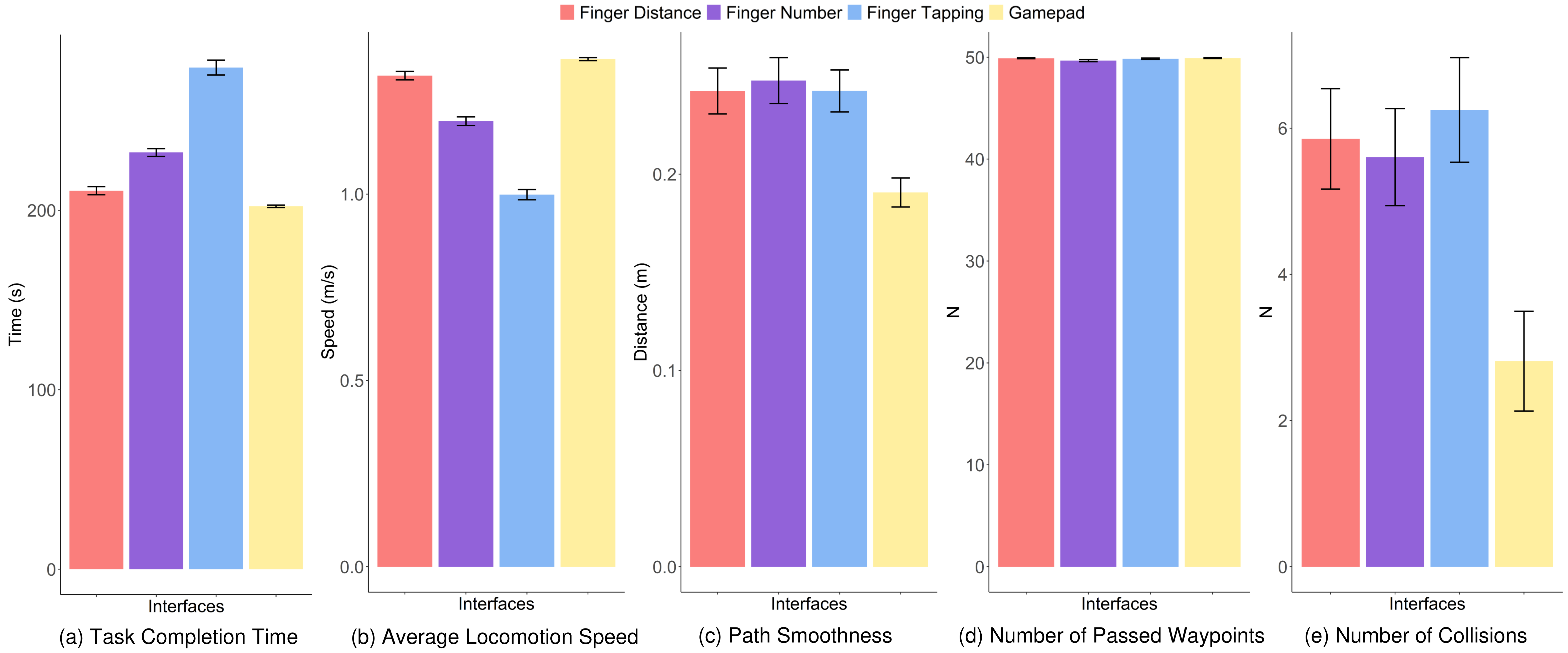}
\caption{Objective factors of Experiment 2 (bar plot convention is as in Figure~\ref{fig:ball_errorbar}).}
\label{fig:objective_door}
\end{figure*}

\begin{table*}[!ht]%
\centering
\caption{Results of Statistical Analyses on Subjective Factors of Experiment 2}
\label{tab:door_subjective}
\includegraphics[width=0.85\textwidth]{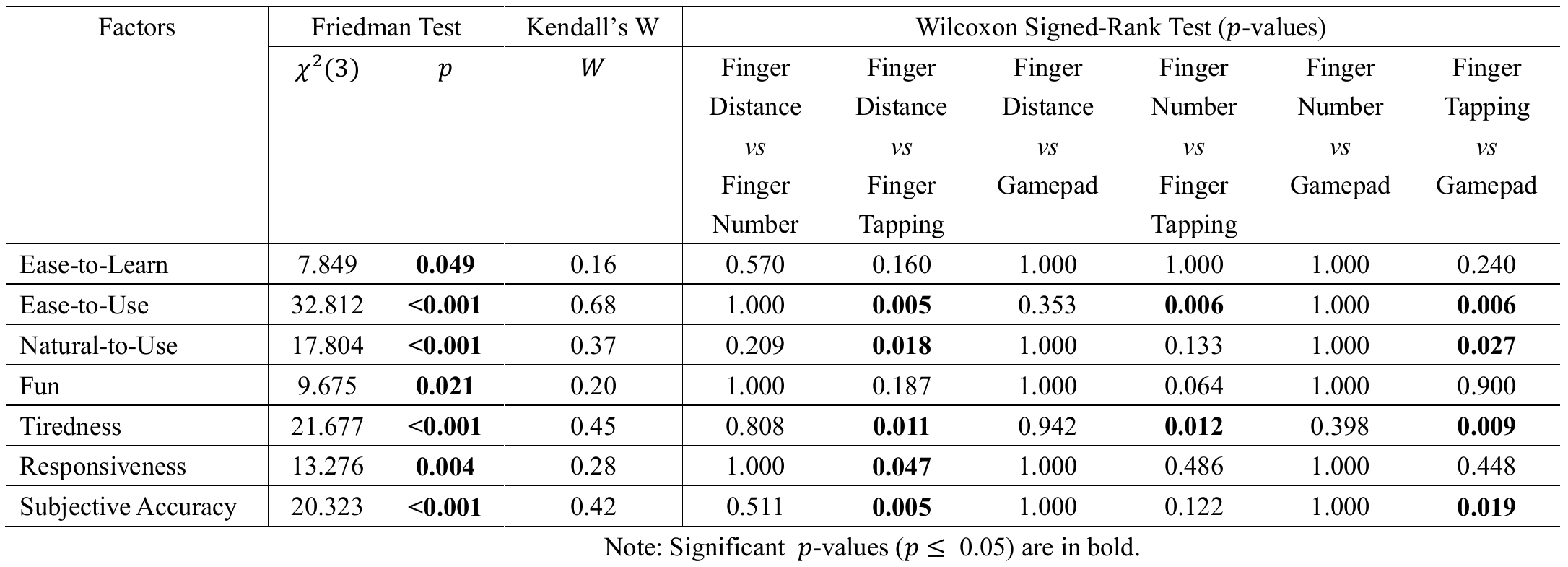}
\end{table*}

\begin{table*}[!ht]%
\centering
\caption{Results of Statistical Analyses on Objective Factors of Experiment 2}
\label{tab:door_objective}
\includegraphics[width=0.6\textwidth]{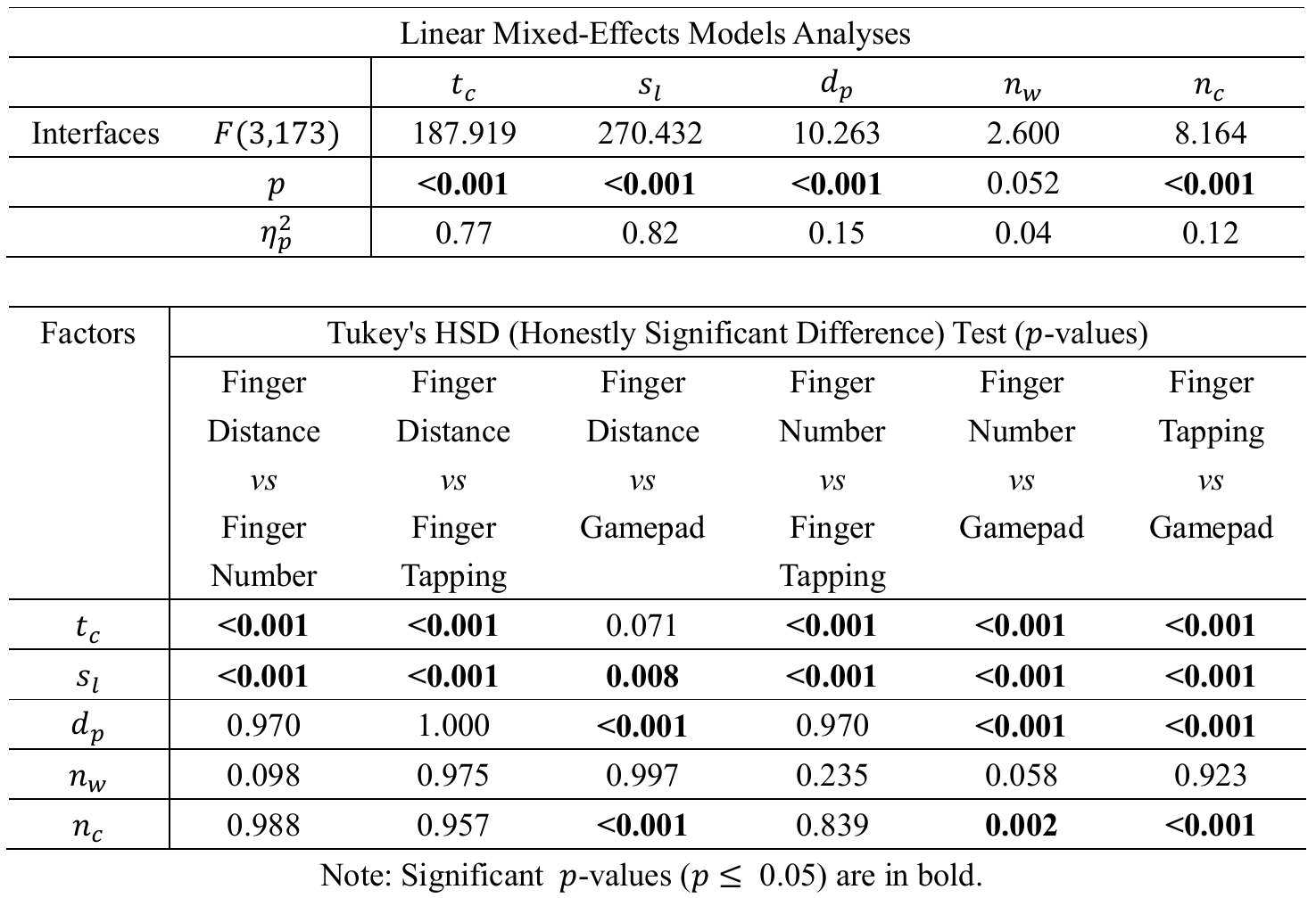}
\end{table*}

\subsubsection{Results}
The protocols for analysing subjective data and objective data in Experiment 2 were as in Experiment 1. The independent factor was the interfaces and the dependent factors for subjective analyses were the factors given in the user interface questionnaire. The dependent factors for objective analyses were defined in Section~\ref{sec:exp2_metrics}. Results of the statistical analyses on both subjective data and objective data are in Table~\ref{tab:door_subjective} and Table~\ref{tab:door_objective}, respectively.

Analyses on subjective factors using the Friedman test showed that there were significant effects on all seven factors: ease-to-learn ($p=0.049$), ease-to-use ($p<0.001$), natural-to-use ($p<0.001$), fun ($p=0.021$), tiredness ($p<0.001$), responsiveness ($p=0.004$) and subjective accuracy ($p<0.001$). However, further analyses using the Wilcoxon signed-rank test with the Bonferroni correction showed that there were no significant differences among four interfaces in terms of ease-to-learn and fun, indicating the effect of interfaces on these factors were weak. Finger Distance, Finger Number and Gamepad were found to be significantly easier to use compared to Finger Tapping, which was consistent with the result in Experiment 1. In terms of natural-to-use, Finger Distance and Gamepad were found to be significantly more natural compared to Finger Tapping but no significant difference was found among Finger Distance, Finger Number and Gamepad, indicating Finger Distance and Gamepad were more intuitive interfaces. Finger Distance, Finger Number and Gamepad were found to be significantly less tiring compared to Finger Tapping. As discussed in Experiment 1, this was mainly due to the repetitive motion of Finger Tapping that resulted in tiredness. Finger Distance was found to be significantly more responsive compared to Finger Tapping as this interface allowed fine control of speed through distance between the thumb and the index of a tracked hand. Finally, Finger Distance and Gamepad were significantly more subjectively accurate than Finger Tapping. The distribution of the user responses in Experiment 2 is shown in Figure~\ref{fig:door_horizontal}. We found that Gamepad still received most ratings of strongly agree, excluding tiredness, compared to other interfaces. Finger Distance and Finger Number had similar number of ratings of strongly agree while Finger Tapping had the lowest.

Objective data analysed using the linear mixed-effects models analyses showed that there were significant effects on parameters, including task completion time $t_c$ ($p<0.001$), average locomotion speed $s_l$ ($p<0.001$), path smoothness $d_p$ ($p<0.001$) and the number of collisions $n_c$ ($p<0.001$) except the number of successfully passed waypoints $n_w$ ($p=0.052$). Post-hoc analyses performed using the Tukey's HSD test combined with examination of Figure~\ref{fig:door_errorbar} showed that Finger Distance and Gamepad had similar task completion time $t_c$ and they were significantly faster compared to other interfaces. Finger Number had a longer task completion time $t_c$ and Finger Tapping was significantly worse compared to other interfaces. In terms of average locomotion speed $s_l$, Gamepad was significantly faster than other interfaces, with Finger Distance being the second, Finger Number being the third and Finger Tapping the last. Locomotion using Gamepad resulted in a significantly smoother path $d_p$ compared to three other interfaces and there was no statistical difference among other three interfaces in terms of path smoothness $d_p$. A similar significant effect was found on the number of collisions $n_c$, in which locomotion using Gamepad had significantly fewer collisions with frames of wooden gates compared to other interfaces. Finger Distance, Finger Number and Finger Tapping had similar performance in terms of the number of collisions $n_c$. A probable reason was that all three gestures used the same direction control mechanism, different speed control mechanisms alone would not impact this factor. There was no significant effect found on the number of successfully passed waypoints as this parameter was a coarse parameter and it was not sensitive to different interfaces. These results showed that the Finger Distance gesture was very efficient in waypoints navigation. The second experiment is a more general case for virtual locomotion with direction control enabled compared to Experiment 1. Therefore, the implications of the results from Experiment 2 are more important and meaningful compared to Experiment 1. Given the variable time durations resulting from using four different interfaces to complete the task and differences in performance of individual participants, the time duration to finish an entire experimental session was approximately 1 h, with the time to fill questionnaires included.

\subsubsection{Summary}
Experiment 2 found that participants had similar user preference on the Finger Distance gesture, the Finger Number gesture and the Gamepad interface while the Finger Tapping gesture was the least preferred. In addition, the performance of the Finger Distance gesture was close to that of the Gamepad interface on the waypoints navigation task. The Finger Number gesture used slightly longer time in completing the task and its average locomotion speed was also lower compared to that of the Finger Distance gesture and the Gamepad interface. The Finger Tapping gesture was the slowest in terms of task completion time and average locomotion speed.

\section{Discussion}
In our study, we found that the Finger Distance gesture had close performance and user preference to the gamepad interface. Our explanation was that the Finger Distance gesture allowed more precise control of walking speed through fine control of the distance between the fingertips of the thumb and the index of a tracked hand compared to two other hand gesture interfaces. The Finger Number gesture was slightly worse on the waypoint navigation task compared to the Finger Distance gesture but it performed better on a few factors in the target pursuit task. A problem with the Finger Number gesture was that the interface gave discrete speed values so it had difficulty in enabling precise control of locomotion speed. It would also have problems in tracking target with constant decimal speed values if this were used as an experimental scenario. However, the Finger Number gesture is still an intuitive technique. It controls locomotion speed based on the number of extended fingers, which is easy for people to learn. The Finger Tapping gesture had the latency problem. Controlling walking speed can only be performed after detecting two consecutive peaks. There was a latency in detecting the peak following a previously detected peak, which made the interface not very responsive in terms of speed control. We expect that by modifying the algorithm to allow speed control within two consecutive peaks (\textit{i.e.} within-step speed control as in the GUD-WIP technique \citep{wendt_gud_2010}), the modified Finger Tapping gesture may perform better than the current method. However, tiredness coming from repetitive tapping motion of finger remains a problem that cannot be easily solved. 

Recently, several commercial VR first person shooter games have appeared. Such games often rely on touch controllers for players to locomote and perform actions in virtual environments with large open space. Such games could offer players an additional choice to use hand gestures for walking and performing other actions in games. Our results showed that the Finger Distance gesture and the Finger Number gesture have such potentials. Another utility for hand gesture interfaces is navigation in projected virtual environments, such as CAVEs \citep{cruz-neira_surround-screen_1993} and the WISE (Wide Immersive Stereo Environment, \citep{zhao_role_2020}). Such virtual environments are often used for architectural inspection or visualisation of large open environments so hand gestures should be potentially useful to such scenarios.

Our study included only one method for direction control, that is, rotating the left palm around the $y$-axis in the Leap Motion sensor's coordinate system. For future work, it is necessary to test the utility of the speed control gestures in combination with different direction control methods proposed by other researchers. These direction control methods include using pointing direction of a single finger \citep{caggianese_freehand-steering_2020} and head orientation \citep{caggianese_freehand-steering_2020,cardoso_comparison_2016}, \textit{etc}. Furthermore, the present methods in our study focused on using finger motions to control locomotion speed. Further research is needed to compare finger motion gestures to static hand gestures \citep{cisse_user_2020} in terms of their usability.

One limitation of our study was that we did not use VR headsets for experiments as sharing headsets among participants is a health concern \citep{steed_evaluating_2020}. We expect this problem to be solved by applying a sanitization procedure for VR headsets once it is established. Other options include recruiting users who have their own VR hardware or establishing a pool of users with funded VR hardware but many challenges remain \citep{steed_evaluating_2020}. We plan to re-conduct experiments using VR headsets in future and compare the results to the present study, which used a curved monitor for virtual scene presentation. Nevertheless, the proposed interfaces should be easily adapted for locomotion in virtual environments using VR headsets in a seated or a standing position, with the Leap Motion sensor mounted on the headsets to track hand movements. This configuration has been demonstrated by previous studies \citep{cardoso_comparison_2016,huang_design_2019}. It is also possible to track hand movements using Oculus Quest 2 without the Leap Motion sensor. However, two issues need to be considered when using a setup with VR headsets. First, when wearing a VR headset, users are unable to see their physical hands. Without showing users' avatar hands in virtual environments, users may struggle to figure out whether their hands are tracked or not. Thus it is necessary to display avatar hands in virtual environments and render necessary visual feedback. For example, when tracking is lost, avatar hands can be rendered static; when hands are successfully tracked, avatar hands can be animated using the actual motions of users' physical hands. Second, the tracking capability of the hand-tracking sensors is limited by their field of view and tracking distance. When users turned their heads to observe virtual environments but their hands are still in their initial position and are out of the view of the tracking sensor, the tracking of users' hands may be lost. There are two potential solutions to this problem. First, better hand-tracking sensors with a larger field of view and tracking distance have to be developed and used. Second, VR designers may consider designing a boundary system that shows the boundary of the tracking volume when users' hands are not in the view of the hand-tracking sensors.

Finally, a few topics can be considered for future research. A very recent framework proposed for evaluating VR locomotion techniques \citep{cannavo_evaluation_2021} offered a new testing tool for virtual locomotion. A direction for future study is to test the present hand gesture interfaces in this framework to study how hand gestures and the gamepad interface perform in more complex locomotion scenarios. Another promising field for further study is to apply these gestures to teleoperation of robots and assess their usability in this type of application. A previous study has investigated the performance of a set of static hand gestures, a gamepad interface and a traditional user interface on a computer desktop for teleoperation of robots \citep{doisy_comparison_2017}. But this study did not evaluate the performance of different hand gestures in teleoperation of robots. Therefore, it is interesting to investigate how different hand gestures perform in such applications and whether hand gesture interfaces improve presence during teleoperation when wearing VR headsets. Lastly, when wearing a VR headset and performing virtual locomotion, simulator sickness \citep{kennedy_simulator_1993} is more likely to occur compared to the desktop setup. But the differences between hand gestures and the gamepad interface of elicited motion sickness are unknown. An interface that leads to severer simulator sickness is less likely to be preferred by users. Thus a study needs to be designed to investigate the differences between hand gestures and the gamepad interface in the degree of elicited simulator sickness.

\section{Conclusion}
In this paper, we presented three hand gestures and their algorithms for virtual locomotion and implemented a gamepad locomotion interface for comparison using the Xbox One controller. Through two virtual locomotion tasks, we systematically compared the performance and user preference of these interfaces using a desktop setup. We showed that the performance and user preference of the Finger Distance gesture were close to that of the gamepad interface. The performance and user preference of the Finger Number gesture were slightly below the Finger Distance gesture. This research provides VR researchers and designers with methods and empirical data to implement locomotion interfaces using hand gestures in VR systems. New VR systems could also include these gesture interfaces and let users select their favourite method during usage. In addition, multi-modal interfaces that integrated different interaction mechanisms (\textit{e.g.} hand gestures, body gestures and voice recognition, \textit{etc.}) for locomotion also can be considered for new VR systems. Such interfaces will provide end-users with more options for interaction in VR. Finally, hand gestures have potentials in their utility for locomotion in mixed reality (MR) applications and teleoperation of robots but further investigations are needed. 



\bibliography{template.bib}

\begin{thebibliography}{33}
\expandafter\ifx\csname natexlab\endcsname\relax\def\natexlab#1{#1}\fi
\providecommand{\url}[1]{\texttt{#1}}
\providecommand{\href}[2]{#2}
\providecommand{\path}[1]{#1}
\providecommand{\DOIprefix}{doi:}
\providecommand{\ArXivprefix}{arXiv:}
\providecommand{\URLprefix}{URL: }
\providecommand{\Pubmedprefix}{pmid:}
\providecommand{\doi}[1]{\href{http://dx.doi.org/#1}{\path{#1}}}
\providecommand{\Pubmed}[1]{\href{pmid:#1}{\path{#1}}}
\providecommand{\bibinfo}[2]{#2}
\ifx\xfnm\relax \def\xfnm[#1]{\unskip,\space#1}\fi
\bibitem[{Bachmann et~al.(2018)Bachmann, Weichert and
  Rinkenauer}]{bachmann_review_2018}
\bibinfo{author}{Bachmann, D.}, \bibinfo{author}{Weichert, F.},
  \bibinfo{author}{Rinkenauer, G.}, \bibinfo{year}{2018}.
\newblock \bibinfo{title}{Review of {Three}-{Dimensional} {Human}-{Computer}
  {Interaction} with {Focus} on the {Leap} {Motion} {Controller}}, in:
  \bibinfo{booktitle}{Sensors}.
\bibitem[{Bozgeyikli et~al.(2016)Bozgeyikli, Raij, Katkoori and
  Dubey}]{bozgeyikli_point_2016}
\bibinfo{author}{Bozgeyikli, E.}, \bibinfo{author}{Raij, A.},
  \bibinfo{author}{Katkoori, S.}, \bibinfo{author}{Dubey, R.},
  \bibinfo{year}{2016}.
\newblock \bibinfo{title}{Point \& {Teleport} {Locomotion} {Technique} for
  {Virtual} {Reality}}, in: \bibinfo{booktitle}{Proceedings of the 2016
  {Annual} {Symposium} on {Computer}-{Human} {Interaction} in {Play}},
  \bibinfo{publisher}{Association for Computing Machinery},
  \bibinfo{address}{New York, NY, USA}. pp. \bibinfo{pages}{205--216}.
\bibitem[{Caggianese et~al.(2020)Caggianese, Capece, Erra, Gallo and
  Rinaldi}]{caggianese_freehand-steering_2020}
\bibinfo{author}{Caggianese, G.}, \bibinfo{author}{Capece, N.},
  \bibinfo{author}{Erra, U.}, \bibinfo{author}{Gallo, L.},
  \bibinfo{author}{Rinaldi, M.}, \bibinfo{year}{2020}.
\newblock \bibinfo{title}{Freehand-{Steering} {Locomotion} {Techniques} for
  {Immersive} {Virtual} {Environments}: {A} {Comparative} {Evaluation}}.
\newblock \bibinfo{journal}{International Journal of Human{\textendash}Computer
  Interaction} \bibinfo{volume}{36}, \bibinfo{pages}{1734--1755}.
\bibitem[{Cannav{\`o} et~al.(2021)Cannav{\`o}, Calandra, Prattic{\`o},
  Gatteschi and Lamberti}]{cannavo_evaluation_2021}
\bibinfo{author}{Cannav{\`o}, A.}, \bibinfo{author}{Calandra, D.},
  \bibinfo{author}{Prattic{\`o}, F.G.}, \bibinfo{author}{Gatteschi, V.},
  \bibinfo{author}{Lamberti, F.}, \bibinfo{year}{2021}.
\newblock \bibinfo{title}{An {Evaluation} {Testbed} for {Locomotion} in
  {Virtual} {Reality}}.
\newblock \bibinfo{journal}{IEEE Transactions on Visualization and Computer
  Graphics} \bibinfo{volume}{27}, \bibinfo{pages}{1871--1889}.
\bibitem[{Cardoso(2016)}]{cardoso_comparison_2016}
\bibinfo{author}{Cardoso, J.C.S.}, \bibinfo{year}{2016}.
\newblock \bibinfo{title}{Comparison of {Gesture}, {Gamepad}, and {Gaze}-based
  {Locomotion} for {VR} {Worlds}}, in: \bibinfo{booktitle}{Proceedings of the
  {22nd} {ACM} {Conference} on {Virtual} {Reality} {Software} and
  {Technology}}, \bibinfo{publisher}{ACM}. pp. \bibinfo{pages}{319--320}.
\bibitem[{Chang and Lin(2011)}]{chang_libsvm:_2011}
\bibinfo{author}{Chang, C.}, \bibinfo{author}{Lin, C.}, \bibinfo{year}{2011}.
\newblock \bibinfo{title}{{LIBSVM}: {A} library for support vector machines}.
\newblock \bibinfo{journal}{ACM Trans. Intell. Syst. Technol.}
  \bibinfo{volume}{2}, \bibinfo{pages}{27:1--27:27}.
\bibitem[{Cisse et~al.(2020)Cisse, Gandhi, Lottridge and
  Amor}]{cisse_user_2020}
\bibinfo{author}{Cisse, K.}, \bibinfo{author}{Gandhi, A.},
  \bibinfo{author}{Lottridge, D.}, \bibinfo{author}{Amor, R.},
  \bibinfo{year}{2020}.
\newblock \bibinfo{title}{User {Elicited} {Hand} {Gestures} for {VR}-based
  {Navigation} of {Architectural} {Designs}}, in: \bibinfo{booktitle}{2020
  {IEEE} {Symposium} on {Visual} {Languages} and {Human}-{Centric}
  {Computing}}, pp. \bibinfo{pages}{1--5}.
\bibitem[{Cruz-Neira et~al.(1993)Cruz-Neira, Sandin and
  DeFanti}]{cruz-neira_surround-screen_1993}
\bibinfo{author}{Cruz-Neira, C.}, \bibinfo{author}{Sandin, D.J.},
  \bibinfo{author}{DeFanti, T.A.}, \bibinfo{year}{1993}.
\newblock \bibinfo{title}{Surround-screen {Projection}-based {Virtual}
  {Reality}: {The} design and implementation of the {CAVE}}, in:
  \bibinfo{booktitle}{Proceedings of the 20th {Annual} {Conference} on
  {Computer} {Graphics} and {Interactive} {Techniques}}, pp.
  \bibinfo{pages}{135--142}.
\bibitem[{Doisy et~al.(2017)Doisy, Ronen and Edan}]{doisy_comparison_2017}
\bibinfo{author}{Doisy, G.}, \bibinfo{author}{Ronen, A.},
  \bibinfo{author}{Edan, Y.}, \bibinfo{year}{2017}.
\newblock \bibinfo{title}{Comparison of three different techniques for camera
  and motion control of a teleoperated robot}.
\newblock \bibinfo{journal}{Applied Ergonomics} \bibinfo{volume}{58},
  \bibinfo{pages}{527--534}.
\bibitem[{Ferracani et~al.(2016)Ferracani, Pezzatini, Bianchini, Biscini and
  Del~Bimbo}]{ferracani_locomotion_2016}
\bibinfo{author}{Ferracani, A.}, \bibinfo{author}{Pezzatini, D.},
  \bibinfo{author}{Bianchini, J.}, \bibinfo{author}{Biscini, G.},
  \bibinfo{author}{Del~Bimbo, A.}, \bibinfo{year}{2016}.
\newblock \bibinfo{title}{Locomotion by {Natural} {Gestures} for {Immersive}
  {Virtual} {Environments}}, in: \bibinfo{booktitle}{Proceedings of the 1st
  {International} {Workshop} on {Multimedia} {Alternate} {Realities}},
  \bibinfo{publisher}{ACM}, \bibinfo{address}{Amsterdam The Netherlands}. pp.
  \bibinfo{pages}{21--24}.
\bibitem[{Huang et~al.(2019)Huang, Harris-adamson, Odell and
  Rempel}]{huang_design_2019}
\bibinfo{author}{Huang, R.}, \bibinfo{author}{Harris-adamson, C.},
  \bibinfo{author}{Odell, D.}, \bibinfo{author}{Rempel, D.},
  \bibinfo{year}{2019}.
\newblock \bibinfo{title}{Design of finger gestures for locomotion in virtual
  reality}.
\newblock \bibinfo{journal}{Virtual Reality \& Intelligent Hardware}
  \bibinfo{volume}{1}, \bibinfo{pages}{1--9}.
\bibitem[{Kennedy et~al.(1993)Kennedy, Lane, Berbaum and
  Lilienthal}]{kennedy_simulator_1993}
\bibinfo{author}{Kennedy, R.S.}, \bibinfo{author}{Lane, N.E.},
  \bibinfo{author}{Berbaum, K.S.}, \bibinfo{author}{Lilienthal, M.G.},
  \bibinfo{year}{1993}.
\newblock \bibinfo{title}{Simulator {Sickness} {Questionnaire}: {An} enhanced
  method for quantifying simulator sickness}.
\newblock \bibinfo{journal}{The International Journal of Aviation Psychology}
  \bibinfo{volume}{3}, \bibinfo{pages}{203--220}.
\bibitem[{Kim et~al.(2008)Kim, Gra{\v c}anin, Matkovi{\'c} and
  Quek}]{kim_finger_2008}
\bibinfo{author}{Kim, J.S.}, \bibinfo{author}{Gra{\v c}anin, D.},
  \bibinfo{author}{Matkovi{\'c}, K.}, \bibinfo{author}{Quek, F.},
  \bibinfo{year}{2008}.
\newblock \bibinfo{title}{Finger {Walking} in {Place} ({FWIP}): {A} {Traveling}
  {Technique} in {Virtual} {Environments}}, in: \bibinfo{booktitle}{Smart
  {Graphics}}. \bibinfo{publisher}{Springer Berlin Heidelberg}. volume
  \bibinfo{volume}{5166}, pp. \bibinfo{pages}{58--69}.
\bibitem[{Kim et~al.(2010)Kim, Gra{\v c}anin, Matkovi{\'c} and
  Quek}]{kim_effects_2010}
\bibinfo{author}{Kim, J.S.}, \bibinfo{author}{Gra{\v c}anin, D.},
  \bibinfo{author}{Matkovi{\'c}, K.}, \bibinfo{author}{Quek, F.},
  \bibinfo{year}{2010}.
\newblock \bibinfo{title}{The {Effects} of {Finger}-{Walking} in {Place}
  ({FWIP}) for {Spatial} {Knowledge} {Acquisition} in {Virtual}
  {Environments}}, in: \bibinfo{booktitle}{Smart {Graphics}},
  \bibinfo{publisher}{Springer}. pp. \bibinfo{pages}{56--67}.
\bibitem[{Lun and Zhao(2015)}]{lun_survey_2015}
\bibinfo{author}{Lun, R.}, \bibinfo{author}{Zhao, W.}, \bibinfo{year}{2015}.
\newblock \bibinfo{title}{A {Survey} of {Applications} and {Human} {Motion}
  {Recognition} with {Microsoft} {Kinect}}.
\newblock \bibinfo{journal}{International Journal of Pattern Recognition and
  Artificial Intelligence} \bibinfo{volume}{29}, \bibinfo{pages}{1555008}.
\bibitem[{Marin et~al.(2016)Marin, Dominio and Zanuttigh}]{marin_hand_2016}
\bibinfo{author}{Marin, G.}, \bibinfo{author}{Dominio, F.},
  \bibinfo{author}{Zanuttigh, P.}, \bibinfo{year}{2016}.
\newblock \bibinfo{title}{Hand gesture recognition with jointly calibrated
  {Leap} {Motion} and depth sensor}.
\newblock \bibinfo{journal}{Multimedia Tools and Applications}
  \bibinfo{volume}{75}, \bibinfo{pages}{14991--15015}.
\bibitem[{Nabiyouni et~al.(2015)Nabiyouni, Saktheeswaran, Bowman and
  Karanth}]{nabiyouni_comparing_2015}
\bibinfo{author}{Nabiyouni, M.}, \bibinfo{author}{Saktheeswaran, A.},
  \bibinfo{author}{Bowman, D.A.}, \bibinfo{author}{Karanth, A.},
  \bibinfo{year}{2015}.
\newblock \bibinfo{title}{Comparing the performance of natural, semi-natural,
  and non-natural locomotion techniques in virtual reality}, in:
  \bibinfo{booktitle}{2015 {IEEE} {Symposium} on {3D} {User} {Interfaces}}, pp.
  \bibinfo{pages}{3--10}.
\bibitem[{Pai et~al.(2018)Pai, Chen, Chan, Isogai, Kimata and
  Kunze}]{pai_pinchmove_2018}
\bibinfo{author}{Pai, Y.S.}, \bibinfo{author}{Chen, Z.}, \bibinfo{author}{Chan,
  L.}, \bibinfo{author}{Isogai, M.}, \bibinfo{author}{Kimata, H.},
  \bibinfo{author}{Kunze, K.}, \bibinfo{year}{2018}.
\newblock \bibinfo{title}{Pinchmove: improved accuracy of user mobility for
  near-field navigation in virtual environments}, in:
  \bibinfo{booktitle}{Proceedings of the 20th {International} {Conference} on
  {Human}-{Computer} {Interaction} with {Mobile} {Devices} and {Services}},
  \bibinfo{publisher}{ACM}, \bibinfo{address}{Barcelona Spain}. pp.
  \bibinfo{pages}{1--11}.
\bibitem[{Rantala et~al.(2021)Rantala, Kangas, Koskinen, Nukarinen and
  Raisamo}]{rantala_comparison_2021}
\bibinfo{author}{Rantala, J.}, \bibinfo{author}{Kangas, J.},
  \bibinfo{author}{Koskinen, O.}, \bibinfo{author}{Nukarinen, T.},
  \bibinfo{author}{Raisamo, R.}, \bibinfo{year}{2021}.
\newblock \bibinfo{title}{Comparison of {Controller}-{Based} {Locomotion}
  {Techniques} for {Visual} {Observation} in {Virtual} {Reality}}.
\newblock \bibinfo{journal}{Multimodal Technologies and Interaction}
  \bibinfo{volume}{5}, \bibinfo{pages}{31}.
\bibitem[{Sarupuri et~al.(2017a)Sarupuri, Hoermann, Steinicke and
  Lindeman}]{sarupuri_triggerwalking_2017}
\bibinfo{author}{Sarupuri, B.}, \bibinfo{author}{Hoermann, S.},
  \bibinfo{author}{Steinicke, F.}, \bibinfo{author}{Lindeman, R.W.},
  \bibinfo{year}{2017}a.
\newblock \bibinfo{title}{Triggerwalking: a biomechanically-inspired locomotion
  user interface for efficient realistic virtual walking}, in:
  \bibinfo{booktitle}{Proceedings of the 5th {Symposium} on {Spatial} {User}
  {Interaction}}, \bibinfo{publisher}{ACM}, \bibinfo{address}{Brighton United
  Kingdom}. pp. \bibinfo{pages}{138--147}.
\bibitem[{Sarupuri et~al.(2017b)Sarupuri, Hoermann, Whitton and
  Lindeman}]{sarupuri_evaluating_2017}
\bibinfo{author}{Sarupuri, B.}, \bibinfo{author}{Hoermann, S.},
  \bibinfo{author}{Whitton, M.C.}, \bibinfo{author}{Lindeman, R.W.},
  \bibinfo{year}{2017}b.
\newblock \bibinfo{title}{Evaluating and {Comparing} {Game}-controller based
  {Virtual} {Locomotion} {Techniques}}.
\newblock \bibinfo{journal}{ICAT-EGVE 2017 - International Conference on
  Artificial Reality and Telexistence and Eurographics Symposium on Virtual
  Environments} , \bibinfo{pages}{7 pages}.
\bibitem[{Sch{\"a}fer et~al.(2021)Sch{\"a}fer, Reis and
  Stricker}]{schafer_controlling_2021}
\bibinfo{author}{Sch{\"a}fer, A.}, \bibinfo{author}{Reis, G.},
  \bibinfo{author}{Stricker, D.}, \bibinfo{year}{2021}.
\newblock \bibinfo{title}{Controlling {Teleportation}-{Based} {Locomotion} in
  {Virtual} {Reality} with {Hand} {Gestures}: {A} {Comparative} {Evaluation} of
  {Two}-{Handed} and {One}-{Handed} {Techniques}}.
\newblock \bibinfo{journal}{Electronics} \bibinfo{volume}{10},
  \bibinfo{pages}{715}.
\bibitem[{Steed et~al.(2020)Steed, Ortega, Williams, Kruijff, Stuerzlinger,
  Batmaz, Won, Rosenberg, Simeone and Hayes}]{steed_evaluating_2020}
\bibinfo{author}{Steed, A.}, \bibinfo{author}{Ortega, F.R.},
  \bibinfo{author}{Williams, A.S.}, \bibinfo{author}{Kruijff, E.},
  \bibinfo{author}{Stuerzlinger, W.}, \bibinfo{author}{Batmaz, A.U.},
  \bibinfo{author}{Won, A.S.}, \bibinfo{author}{Rosenberg, E.S.},
  \bibinfo{author}{Simeone, A.L.}, \bibinfo{author}{Hayes, A.},
  \bibinfo{year}{2020}.
\newblock \bibinfo{title}{Evaluating immersive experiences during {Covid}-19
  and beyond}.
\newblock \bibinfo{journal}{Interactions} \bibinfo{volume}{27},
  \bibinfo{pages}{62--67}.
\bibitem[{Teknomo(2002)}]{teknomo_microscopic_2002}
\bibinfo{author}{Teknomo, K.}, \bibinfo{year}{2002}.
\newblock \bibinfo{title}{Microscopic {Pedestrian} {Flow} {Characteristics}:
  {Development} of an {Image} {Processing} {Data} {Collection} and {Simulation}
  {Model}}.
\newblock Ph.D. thesis. Tohoku University.
\bibitem[{Tomberlin et~al.(2017)Tomberlin, Tahai and
  Pietroszek}]{tomberlin_gauntlet_2017}
\bibinfo{author}{Tomberlin, M.}, \bibinfo{author}{Tahai, L.},
  \bibinfo{author}{Pietroszek, K.}, \bibinfo{year}{2017}.
\newblock \bibinfo{title}{Gauntlet: {Travel} technique for immersive
  environments using non-dominant hand}, in: \bibinfo{booktitle}{2017 {IEEE}
  {Virtual} {Reality} ({VR})}, \bibinfo{publisher}{IEEE}, \bibinfo{address}{Los
  Angeles, CA, USA}. pp. \bibinfo{pages}{299--300}.
\bibitem[{Tregillus and Folmer(2016)}]{tregillus_vr-step:_2016}
\bibinfo{author}{Tregillus, S.}, \bibinfo{author}{Folmer, E.},
  \bibinfo{year}{2016}.
\newblock \bibinfo{title}{{VR}-{STEP}: {Walking}-in-{Place} using inertial
  sensing for hands free navigation in mobile {VR} environments}, in:
  \bibinfo{booktitle}{Proceedings of the 2016 {CHI} {Conference} on {Human}
  {Factors} in {Computing} {Systems}}, pp. \bibinfo{pages}{1250--1255}.
\bibitem[{Wendt et~al.(2010)Wendt, Whitton and Brooks}]{wendt_gud_2010}
\bibinfo{author}{Wendt, J.D.}, \bibinfo{author}{Whitton, M.C.},
  \bibinfo{author}{Brooks, F.P.}, \bibinfo{year}{2010}.
\newblock \bibinfo{title}{{GUD} {WIP}: {Gait}-{Understanding}-{Driven}
  {Walking}-{In}-{Place}}, in: \bibinfo{booktitle}{2010 {IEEE} {Virtual}
  {Reality} {Conference}}, pp. \bibinfo{pages}{51--58}.
\bibitem[{Yan et~al.(2016)Yan, Lindeman and Dey}]{yan_let_2016}
\bibinfo{author}{Yan, Z.}, \bibinfo{author}{Lindeman, R.W.},
  \bibinfo{author}{Dey, A.}, \bibinfo{year}{2016}.
\newblock \bibinfo{title}{Let your fingers do the walking: {A} unified approach
  for efficient short-, medium-, and long-distance travel in {VR}}, in:
  \bibinfo{booktitle}{2016 {IEEE} {Symposium} on {3D} {User} {Interfaces}}, pp.
  \bibinfo{pages}{27--30}.
\bibitem[{Zhang et~al.(2017)Zhang, Chu, Pan, Ji and Xi}]{zhang_double_2017}
\bibinfo{author}{Zhang, F.}, \bibinfo{author}{Chu, S.}, \bibinfo{author}{Pan,
  R.}, \bibinfo{author}{Ji, N.}, \bibinfo{author}{Xi, L.},
  \bibinfo{year}{2017}.
\newblock \bibinfo{title}{Double hand-gesture interaction for walk-through in
  {VR} environment}, in: \bibinfo{booktitle}{2017 {IEEE}/{ACIS} 16th
  {International} {Conference} on {Computer} and {Information} {Science}
  ({ICIS})}, \bibinfo{publisher}{IEEE}, \bibinfo{address}{Wuhan, China}. pp.
  \bibinfo{pages}{539--544}.
\bibitem[{Zhao and Allison(2018)}]{zhao_learning_2016}
\bibinfo{author}{Zhao, J.}, \bibinfo{author}{Allison, R.S.},
  \bibinfo{year}{2018}.
\newblock \bibinfo{title}{Learning gait parameters for locomotion in virtual
  reality systems}, in: \bibinfo{booktitle}{Understanding {Human} {Activities}
  {Through} {3D} {Sensors}}, \bibinfo{publisher}{Springer, Cham}. pp.
  \bibinfo{pages}{59--73}.
\bibitem[{Zhao and Allison(2020)}]{zhao_comparing_2019}
\bibinfo{author}{Zhao, J.}, \bibinfo{author}{Allison, R.S.},
  \bibinfo{year}{2020}.
\newblock \bibinfo{title}{Comparing head gesture, hand gesture and gamepad
  interfaces for answering {Yes}/{No} questions in virtual environments}.
\newblock \bibinfo{journal}{Virtual Reality} \bibinfo{volume}{24},
  \bibinfo{pages}{515--524}.
\bibitem[{Zhao and Allison(2021)}]{zhao_role_2020}
\bibinfo{author}{Zhao, J.}, \bibinfo{author}{Allison, R.S.},
  \bibinfo{year}{2021}.
\newblock \bibinfo{title}{The {Role} of {Binocular} {Vision} in {Avoiding}
  {Virtual} {Obstacles} {While} {Walking}}.
\newblock \bibinfo{journal}{IEEE Transactions on Visualization and Computer
  Graphics} \bibinfo{volume}{27}, \bibinfo{pages}{3277--3288}.
\bibitem[{Zhao et~al.(2018)Zhao, Allison, Vinnikov and
  Jennings}]{zhao_effects_2018}
\bibinfo{author}{Zhao, J.}, \bibinfo{author}{Allison, R.S.},
  \bibinfo{author}{Vinnikov, M.}, \bibinfo{author}{Jennings, S.},
  \bibinfo{year}{2018}.
\newblock \bibinfo{title}{The {Effects} of {Visual} and {Control} {Latency} on
  {Piloting} a {Quadcopter} {Using} a {Head}-{Mounted} {Display}}, in:
  \bibinfo{booktitle}{2018 {IEEE} {International} {Conference} on {Systems},
  {Man}, and {Cybernetics}}, pp. \bibinfo{pages}{2972--2979}.

\end{thebibliography}

\end{document}